\documentclass[a4paper,11pt]{article}
\pdfoutput=1 

\usepackage{jheppub} 

\usepackage[T1]{fontenc} 

\usepackage{Settings}
\usepackage{simpler-wick}

\usepackage{graphicx}
\usepackage{dcolumn}
\usepackage{bm}

\newcommand{\nn}{\nonumber}

\newcommand{\cg}{g_{\text{\tiny YM}}}
\renewcommand{\vec}[1]{\mathbf{#1}}

\makeatletter
\newlength{\apb@width}
\newcommand{\autoparbox}[2][c]{\settowidth{\apb@width}{#2}\parbox[#1]{\apb@width}{#2}}
\newcommand{\includegraphicsbox}[2][]{\autoparbox{\includegraphics[#1]{#2}}}
\makeatother

\makeatletter
\def\mr@ignsp#1 {\ifx\:#1\@empty\else #1\expandafter\mr@ignsp\fi}%
\newcommand{\multiref}[1]{\begingroup
	\xdef\mr@no@sparg{\expandafter\mr@ignsp#1 \: }%
	\def\mr@comma{}%
	\@for\mr@refs:=\mr@no@sparg\do{\mr@comma\def\mr@comma{,}\ref{\mr@refs}}%
	\endgroup}
\makeatother

\newcommand{\hypref}[2]{\ifx\href\asklfhas #2\else\href{#1}{#2}\fi}

\newcommand{\secref}[1]{Sec.~\multiref{#1}}

\newcommand{\figref}[1]{Fig.~\multiref{#1}}
\renewcommand{\eqref}[1]{(\multiref{#1})}

\def\<{\begin{eqnarray}}
\def\>{\end{eqnarray}}

\usepackage{Settings}

\begin{document}

\title{Asymptotic Charges and Coherent States in QCD}

\author[a]{Riccardo Gonzo} \author[a]{, Tristan Mc Loughlin} \author[b]{, Diego Medrano} \author[a]{and Anne Spiering}

\affiliation[a]{School of Mathematics \& Hamilton Mathematics Institute,\\
	 Trinity College Dublin, Ireland}
\affiliation[b]{Instituto de F{\'i}sica Te{\'o}rica UAM/CSIC \& Universidad Aut{\'o}noma de Madrid,\\
	 C/Nicol{\'a}s Cabrera 15, E-28049 Madrid, Spain }

\emailAdd{gonzo@maths.tcd.ie}
\emailAdd{tristan@maths.tcd.ie}
\emailAdd{d.medrano@csic.es}
\emailAdd{spiering@maths.tcd.ie}

\date{\today}

\abstract{We study the connection between asymptotic symmetries in non-Abelian gauge theories and the generalised coherent states following from the application to QCD of the Faddeev-Kulish approach to asymptotic dynamics. We compute the large gauge transformation properties of the soft evolution operators and use this to define the quantum corrected, non-linear contribution to the asymptotic charges. We then compute the  leading IR-divergent part of the one-loop correction to matrix elements of the charges and show that, with a specific ordering of soft limits, the  asymptotic charges continue to be conserved at this order.}

\maketitle


\section{\label{sec:intro}Introduction}

The study of the infrared behaviour of gauge theory scattering amplitudes has a long history and in the case of QED has essentially been understood since the work of Bloch and Nordsieck \cite{Bloch:1937pw}, though it has been refined over the years \cite{Yennie:1961ad, Kinoshita:1962ur, Lee:1964is,  Weinberg:1965nx}. 
The standard approach involves the computation of amplitudes which are formally singular - they in fact vanish after exponentiation of the perturbative divergences. One then focuses on inclusive quantities involving arbitrary numbers of real soft-photons which cancel the IR-divergences from virtual photons in loops.  An alternative approach is to directly formulate infrared-finite S-matrix elements by choosing appropriate asymptotic states. For QED this approach where the asymptotic states are not eigenstates of the photon number operator but rather have the form of coherent states, was, starting from the work of Chung \cite{Chung:1965zza} \footnote{The idea of removing the IR-divergences from S-matrix elements by using appropriately dressed asymptotic states was independently considered in  \cite{Greco:1967zza}. We are grateful to Mario Greco for bringing this reference to our attention.}, developed by a number of authors, e.g.\ \cite{Kibble:1968sfb, Kibble:1969ip, Zwanziger:1973if, Schweber:1973qa}. Faddeev and Kulish, building on the work of Dollard for the Coulomb problem in non-relativistic quantum mechanics \cite{dollard1964asymptotic}, related the structure of these coherent states to the form of the large-time Hamiltonian \cite{Kulish:1970ut,Horan:1999ba}. The approach was partially extended to the much more complicated case of non-Abelian gauge theory \cite{GRECO1978282, Dahmen:1980nw, Butler:1978, Nelson:1980yt, FRENKEL1982}, \cite{CATANI1985, CATANI1986,CATANI1987, CIAFALONI1985, Ciafaloni1989, DELDUCA1989} and more recently to perturbative gravity \cite{Ware:2013zja}.

The observation that asymptotic conservation laws, which follow from
Noether's second theorem \cite{Noether1918} for large gauge transformations, pave the way to understanding the infrared dynamics of gauge theories \cite{Strominger:2013lka, Strominger:2013jfa} (see \cite{Strominger:2017zoo} for a review and more complete references), has lead to a renewed interest in the study of coherent state operators and soft-dressing more generally \cite{Mirbabayi:2016axw, Gabai:2016kuf, Kapec:2017tkm,  Choi:2017bna, Choi:2017ylo, Gomez:2017rau, Carney:2018ygh, Neuenfeld:2018fdw, Hirai:2019gio}.   
It has already been demonstrated that for QED \cite{Kapec:2017tkm} and perturbative gravity \cite{Choi:2017ylo} the   coherent states relevant to the construction of an infrared finite S-matrix follow from the symmetry of asymptotic charges. 
The existence of an infinite dimensional symmetry group has lead to the interpretation of the QED vacuum as being degenerate and that  scattering processes are accompanied by a shift in the vacua. Infrared divergences due to massless particles which result in the vanishing of S-matrix elements are thus connected with the ``wrong" choice of the in- and out-vacua, and a cure can be sought in a systematic way by considering the asymptotic charges. An analogous statement can be made for perturbative gravity, using BMS supertranslation charges to find suitable asymptotic states for an infrared-finite gravity S-matrix.

The corresponding understanding of QCD infrared dynamics is significantly less complete. Compared to QED, the natural complication that arises is that gauge bosons self-interact in a non-trivial way. The persistence of these non-trivial self-interactions at early and late times is central to the behaviour of the infrared properties of QCD\cite{Pate:2017vwa}. It is our aim here to understand the connection between asymptotic symmetries in non-Abelian gauge theories and generalized IR leading coherent states. We will follow the approach of \ Catani, \ Ciafaloni and \ Marchesini,  \cite{CATANI1985, CATANI1986, CATANI1987, CIAFALONI1985, Ciafaloni1989} (see \cite{book:Mueller} for an introduction) which uses energy ordering in each interaction to systematically organise the divergences due to soft gluons. Non-Abelian gauge theories of course also have collinear divergences which have been treated in the coherent operator approach \cite{DelDuca:1989jt}, see also \cite{Forde:2003jt} and the recent interesting work \cite{Hannesdottir:2019rqq, Hannesdottir:2019opa}.  We will focus on the leading IR-divergence, though the method can be extended in principle to include sub-leading divergences. To be slightly more precise, when we extend the tree-level considerations to one-loop we will focus on the IR double pole terms arising from simultaneously soft and collinear regions of parameter space.

In the following a central goal is the definition of asymptotic charges at loop level. As we discuss below we start from the linear part of the charge that arises in the classical construction and which can be understood to act on Fock, or bare, asymptotic states at null-infinity. However, as the S-matrix is ill-defined for such states we must introduce the soft-evolution operators generated by the QCD Hamiltonian which is used to define dressed states. We then consider the transformation properties of these soft-evolution operators under transformations generated by the asymptotic charges and we show how the soft-evolution operators can be viewed as generating the non-linear part of the charge from the linear part. This implies the existence of an infinite sequence of quantum corrections to the non-linear charge.  This charge can then be used to define a Ward identity for scattering amplitudes. 
The main result of this paper is the computation, at leading IR divergence, of the one-loop corrections to the Ward identity of the asymptotic charge which is related to the soft gluon theorem. We demonstrate explicitly that the contribution from the soft gluon dressing factor cancels the contribution from the one-loop soft factor in QCD (\cite{Catani:2000pi, Bierenbaum:2011gg}), showing that the asymptotic charge produces a vacuum that is orthogonal to all scattering states built on the original vacuum as in \cite{Kapec:2017tkm}. We will see that this result depends on the precise prescription for the order of soft limits in the definition of the charge and dressing factor which is analogous to the order of limits discussed in \cite{Bern:2014oka,Cachazo:2014dia}. There are two order-of-limit choices we must make, for the first, in the definition of the charge we take a order of limits used in \cite{Bern:2014oka}, which corresponds to the standard choice in dimensionally-regularized soft limits. This choice is motivated, as in \cite{Bern:2014oka}, by its use in problems such as the computation of physical cross-sections. For the second choice, in the definition of the dressed states, we compute the result using both orderings for comparison. As we discuss in \secref{sec:disc} it is however natural to choose the prescription which preserves, where possible, the symmetries of the theory. We will see that there is indeed a particular prescription that preserves the asymptotic symmetries to the order we compute, as we find that with this choice the Ward identity receives no corrections at one-loop and leading IR-divergence.

\paragraph{Preliminaries}

Perturbative computations are relevant in QCD as they are related to experimental observables due to two important properties: asymptotic freedom and factorisation. Confinement naturally sets a scale $\Lambda_{\text{QCD}}$ such that for partons to be well-defined objects we require the existence of a lower cutoff 
on the lowest momentum transfer in a given process. Perturbative quantities are then related to physical cross-sections by convolution with non-pertubative but universal objects, e.g.\ parton distribution functions.
 In this work we will initially consider asymptotic states formed by acting on the Fock vacuum with creation/annihilation operators for the hard incoming or outgoing partons (these can be either massless gluons or massive quarks)
\begin{align}
\prod_{\ell}  b^\dagger_{\alpha_\ell, \sigma_\ell}(p_\ell)\ket{0}=\ket{ \{p_\ell, \alpha_\ell, \sigma_\ell\}}
\end{align}
which are labelled by momenta $p_\ell$,  colour indices $\alpha_\ell$ (corresponding to the fundamental representation for quarks and the adjoint for gluons) and helicity indices $\sigma_\ell$ as appropriate.

We will consider $n$-particle IR divergent S-matrix elements between such asymptotic states
\begin{align}
\label{eq:tree_soft}
\hspace{-30pt}\mathcal{M}_n(\{p_\ell,\alpha_\ell,\sigma_\ell\})=&\langle 0| \left(\prod_{\ell \in \text{out}}  b_{\alpha_\ell,\sigma_\ell }(p_\ell)\right)S\left(\prod_{\ell \in \text{in}}  b^\dagger_{\alpha_\ell,\sigma_\ell}(p_\ell)\right) |0\rangle\nn\\
=& \cg^{n-2}\sum_{L=0}^\infty \cg^{2L}\mathcal{M}^{(L)}_n(\{p_\ell,\alpha_\ell,\sigma_\ell\})
\end{align}
which give the usual perturbative scattering amplitudes. The behaviour of such amplitudes as individual gluons become soft and the relation to the asymptotic charge plays a key role in our work. We will properly introduce and define the linearized asymptotic charge, $Q^{\text{lin}}_{\epsilon}$,  in subsequent sections however for now it is sufficient to state that it involves soft gluon creation/annihilation operators $a^{a\dagger}_\sigma(\omega_q)$/$a^a_\sigma(\omega_q)$ with vanishing gluon energy $\omega_q\simeq 0$ and has matrix elements of the form
\begin{align}
\langle \text{out} | Q^{\text{lin}}_\epsilon S |\text{in}\rangle \sim \langle \text{out} | \lim_{\omega_q \to 0} \omega_q a^a_{\sigma}(\omega_q) S |\text{in}\rangle  
\label{eqn:Qcharge_lin}
\end{align}
with a careful interpretation of the limit.
At tree level it is known that the limit in the definition of the charge can be understood as
\begin{align}
\langle \text{out} | \lim_{\omega_q \to 0} \omega_q a^a_{\sigma}(\omega_q) S |\text{in} \rangle =& \lim_{\omega_q \to 0} \omega_q \langle\text{out} | a^a_{\sigma}(\omega_q) S |\text{in}\rangle  \nonumber\\ =&\lim_{\omega_q \to 0} \omega_q \mathcal{M}^{(0)}_{n+1}(\{q, a, \sigma\})~.
\end{align}
These terms can be computed using the tree-level soft-gluon theorem 
\begin{align}
\lim_{\omega_q \to 0} \mathcal{M}^{(0)}_{n+1}(\{q, a,\sigma\},\{p_\ell\})= g_{\text{\tiny YM}}  J^{(0)a}_\sigma(q) \mathcal{M}^{(0)}_n(\{p_\ell\})
\end{align}
with the soft-current given by
\begin{align}
\label{eq:treesoftcurr}
J^{(0)a}_\sigma(q)=\Big[ \sum_{\ell\in \text{out} }  \frac{p_\ell \cdot { \varepsilon}^\sigma (q)}{p_\ell\cdot q}  t_\ell^{a} -  \sum_{\ell\in \text{in} }  \frac{p_\ell \cdot {\varepsilon}^\sigma (q)}{p_\ell\cdot q}  t_\ell^{a} \Big]~,
\end{align}
where the soft gluon of momentum $q$, colour $a$ and helicity $\sigma$ is taken to be outgoing. In this expression, and similar expressions below, as the limit does not strictly exist the notation $\lim_{\omega_q \to 0}$ should be understood as referring to the leading term in an  expansion in small $\omega_q$. If the gluon was incoming, there would be an overall minus sign and the corresponding polarisation vector would be $\bar\varepsilon^\sigma$. 

 At loop level the issue of the soft limit is more subtle: after computing the perturbative terms defining the matrix elements one may attempt to take the soft limit at the level of the integrands before performing loop integrations or alternatively one may keep $\omega_q$ finite and take the limit only after performing the loop integrations. It is known, from the case of subleading IR behaviour of graviton amplitudes \cite{Bern:2014oka,Cachazo:2014dia,He:2014bga}, that the order has important consequences for the interpretation of quantum corrections to the Ward identities for asymptotic symmetries. In this work for the definition of matrix elements of the asymptotic charge  we always take the soft limit after the computation of matrix elements. That is we will define a regularised charge $Q^\text{lin}_{\epsilon}(\omega_q)$ and then define
\begin{align}
\langle \text{out}| Q^{\text{lin}}_{\epsilon}(0) S |\text{in} \rangle := \lim_{\omega_q \to 0}  \langle \text{out} | Q^{\text{lin}}_{\epsilon}(\omega_q) S |\text{in} \rangle 
\end{align}
so that for the insertion of the soft gluon operator we use the usual soft gluon theorem. The regularisation of the charge will be related to that of the soft-limits of amplitudes. When we discuss the soft-evolution operators we will follow \cite{CIAFALONI1985} and use an energy cut-off. However for explicit computations we will make use of the one-loop soft-limits which were computed using dimensional regularisation, with parameter $\hat{\epsilon}=(d-4)/2$, in \cite{Bierenbaum:2011gg} (see also \cite{Kunszt:1994np,Catani:1996jh,Catani:1996vz,Catani:1998bh,Catani:1999ss,Catani:2000pi,Li:2013lsa} for an earlier work in the case where the quarks are massless) and can be written as
\begin{align}
\label{eq:1loopsoft}
\lim_{\omega_q\to 0} &\mathcal{M}^{(1)}_{n+1}(\{q, a, \sigma\}, \{p_\ell\})=
g_{\text{\tiny YM}} J^{(0)a}_\sigma(q) \mathcal{M}^{(1)}_n( \{p_\ell\})
+g_{\text{\tiny YM}}^3 J^{(1)a}_\sigma(q) \mathcal{M}^{(0)}_n(\{p_\ell\})~,
\end{align}
where the first term on the right-hand side is the iterated tree result involving the tree-level soft current \eqref{eq:treesoftcurr} while the second term is due to the one-loop soft current which is, to leading divergence, 
\begin{align}
J^{(1)a}_\sigma(q)=-\frac{C_A}{16\pi^2 \hat{\epsilon}^2}J^{(0)a}_\sigma(q) + \mathcal{O} \left(\frac{1}{\hat{\epsilon}}\right)~,
\end{align}
where $C_A$ is the adjoint quadratic Casimir. It is important to note that this leading double pole is due to both collinear and soft divergences at leading logarithmic accuracy and as we will see the coherent state will deal with both of them at this order, as explained in \cite{CATANI1986,Ciafaloni1989}.

However before computing such matrix elements we first understand in more detail the definition of the asymptotic charge and its expression in terms of free-field operators. In order to do this we must review both the classical construction of the charges, the soft-evolution operators in QCD and the transformation properties of the evolution operators under large gauge transformations. 


\section{Asymptotic Charges for QCD}
\label{sec:charges}

We will be interested in the asymptotic charges of Yang-Mills theory related to large gauge transformations, which are those that are non-vanishing on the boundary of space-time. Quite generally, Noether's second theorem relates the existence of a local symmetry to a two-form $\kappa^{\nu \mu}$ which can be integrated over a codimension-two sphere to define a charge. That is, the local symmetry implies that there exists a conserved current
\begin{align}
j^{\mu}(\epsilon) = S^{\mu}(\epsilon) + \partial_{\nu} \kappa^{\nu \mu}(\epsilon)~,
\end{align}
where the function $\epsilon(x)$ parametrizes local symmetry transformations and the current $S^{\mu}(\epsilon)$ vanishes on-shell
\<
\qquad S^{\mu}(\epsilon) \stackrel{\text{eom}}{=} 0  ~.
\>
If $\epsilon$ is constant, one recovers the usual conserved current that couples to the gauge field. If we consider the integral of $j^{\mu}(\epsilon)$ over a manifold $\Sigma$
\begin{align}
\int_{\Sigma} d \Sigma_{\mu} j^{\mu}(\epsilon) &= \int_{\Sigma} d \Sigma_{\mu} \big[S^{\mu}(\epsilon) + \partial_{\nu} \kappa^{\nu \mu}(\epsilon)\big] \nn\\
&\stackrel{\text{eom}}{=} \int_{\sigma = \partial \Sigma} d \sigma_{\mu \nu} \kappa^{\nu \mu}(\epsilon)~,
\end{align}
this charge is non-vanishing only if the function $\epsilon$ has support at the manifold boundary $\partial \Sigma$, otherwise it is trivially zero.  This is equivalent to the statement that gauge symmetries are not really ``symmetries" in the proper sense. 

Let us review the construction of classical non-Abelian asymptotic charges at $\mathcal{I}^{\pm}$ for QCD in four dimensional Minkowski space using the usual conformal compactification (for details see \cite{Strominger:2013lka, Strominger:2017zoo} and see Appendix \ref{sec:appendixA0} for the notations used in this paper). The two-form, $\kappa$, is given by the 
field strength associated to the gauge field $\mathcal{A}_{\mu} = \mathcal{A}_{\mu}^a T^a$
\begin{align}
\mathcal{F}_{\mu \nu} = \partial_{\mu} \mathcal{A}_{\nu} - \partial_{\nu} \mathcal{A}_{\mu} -i  \cg [\mathcal{A}_{\mu}, \mathcal{A}_{\nu}] 
\end{align}
which obeys the equations of motion
\begin{align}
\nabla^{\mu} \mathcal{F}_{\mu \nu} -i\cg [\mathcal{A}^{\mu}, \mathcal{F}_{\mu \nu}] = \cg  j_{\nu}^{M}~,
\end{align}
where $j_{\nu}^{M}$ is the matter colour current. The classical charge for non-Abelian large gauge transformations with parameter $\epsilon$ is then given as, 
\begin{align}
Q_\epsilon = \int_{\sigma} \ast \text{tr} \left[\epsilon  \mathcal{F} \right]~,
\label{eqn:charge_YM}
\end{align}
see for example \cite{Barnich:2001jy, Avery:2015rga} and also \cite{Abbott:1982jh}. 

\begin{figure}
	\centering
	\includegraphicsbox[scale=0.9]{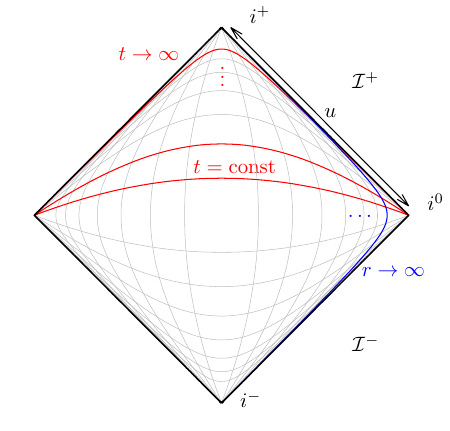}
	\caption{Penrose diagram picture of Minkowski compactification}
	\label{fig:Penrose}
\end{figure}

The relevant surfaces for our asymptotic charges are future and past null infinity, usually denoted $\mathcal{I}^{+}$ and $\mathcal{I}^{-}$ (see \figref{fig:Penrose}).
The natural coordinates for discussing $\mathcal{I}^{+}$ are the so-called retarded Bondi coordinates
\begin{align}
r^2 =& \sum_{i = 1}^3 (x_i)^2~, \qquad u = t - r~, \qquad z = \frac{x_1 + i x_2}{r + x_3}~,
\end{align}
such that the flat-space metric becomes
\begin{align} 
d s^2 =& - d u^2 - 2 d u d r + 2 r^2 \gamma_{z \bar{z}} d z d \bar{z} \quad \text{with} \quad \gamma_{z \bar{z}} = \frac{2}{(1 + z \bar{z})^2}~.
\end{align}
The boundary at $r\to \infty$, $\mathcal{I}^{+}$, is thus parametrized by the coordinates $(u, z, \bar{z})$.
The advanced coordinates, $(r, v=t+r, z, \bar{z})$, are most convenient for $\mathcal{I}^{-}$ and where necessary we will apply the antipodal matching conditions as in \cite{Strominger:2013jfa}.
Furthermore, as it is convenient for making contact with perturbative computations,  relevant for the construction of the coherent states, we will focus on Lorenz gauge in this paper with the gauge fixing condition
\begin{align}
\nabla^{\mu} \mathcal{A}^a_{\mu} = 0~.
\label{eqn:gaugedef}
\end{align}
The same charge can be found in non-covariant gauges such as retarded radial or radiation gauge. In general the asymptotic symmetries may depend on the choice of 
gauge fixing, which can also be seen at the path integral level (see \cite{Avery:2015rga}), but for the leading asymptotic charges one finds the same result for both covariant and physical gauge choices. 
Finally we will be interested in large gauge transformations parametrised by a function $\epsilon = \epsilon^a(z, \bar{z}) T^a$ which labels the transformations at infinity, for example at $\mathcal{I}^{+}$,
\begin{align}
\delta_{\epsilon} A_z(u , z, \bar{z}) = D_z \epsilon(z, \bar{z})~,
\end{align}
where $A_z(u , z, \bar{z})=\lim_{r\to \infty} \mathcal{A}_z(r, u , z, \bar{z})$. 

In retarded Bondi coordinates on the celestial sphere $\mathcal S^2$ defined by the $u\to -\infty$ region of $\mathcal{I}^{+}$, i.e.\ $\mathcal{I}^{+}_-$, the charge \eqref{eqn:charge_YM} becomes 
\begin{equation}
Q_\epsilon =  \int_{\mathcal{I}^{+}_{-}} d^2 z \gamma_{z \bar{z}}~ \text{tr} \left[\epsilon(z, \bar{z}) F_{r u} \right]~,
\label{eq:Q_components_Lorenz}
\end{equation}
where $F_{r u}=\lim_{r\to \infty} r^2 \mathcal{F}_{ru}$. This charge is usually denoted $Q^{+}_\epsilon$ to distinguish it from the corresponding charge $Q^{-}_\epsilon$ on $\mathcal{I}^{-}$. In order to avoid a proliferation of superscripts we will mostly neglect this index and hopefully it is clear from context to which charge we are referring.

 The charge can now be rewritten, assuming the field strengths vanish at $\mathcal I^+_+$ (for details see appendix \ref{sec:appendixA}), as 
\begin{align}
Q_\epsilon = Q_\epsilon^{\text{lin}} + Q^{\text{non-lin}}_\epsilon~,
\end{align}
where
\begin{align}
\label{eq:Q_lin}
Q^{\text{lin}}_\epsilon = \int_{\mathcal{I}^{+}} d^2 z d u~  \epsilon^a(z, \bar{z}) \left[ \partial_u (\partial_z A^a_{\bar{z}} + \partial_{\bar{z}} A^a_z) \right]
\end{align}
and
\begin{align}
Q^{\text{non-lin}}_\epsilon &= \cg \int_{\mathcal{I}^{+}} d^2 z d u  ~\epsilon^a(z, \bar{z})  \left[ f^{a b c} (A^b_z \partial_u A^c_{\bar{z}} - A^c_{\bar{z}} \partial_u A^b_z) +   \gamma_{z \bar{z}} j_u^{a(2)} \right]~.
\label{eq:Q_non-lin}
\end{align}
Here we have adopted the standard splitting of the asymptotic charge into a piece linear in the gauge field (usually called soft) and a piece non-linear in the gauge fields (usually called hard). 
In the construction of asymptotic charges it is usually argued that it is the linearised theory which is used to define the two-form. This is the case in the framework of covariant charges \cite{Barnich:2001jy, Barnich:2003xg} and in a related context was noted by Abbott and Deser \cite{Abbott:1982jh} for Yang-Mills theory. 
In the current context this would correspond to  linearising the theory around the background $A^{a}_z = A^{a}_{\bar z} = 0$ and would thus discard the non-linear gluon piece of the asymptotic charge. However,  
it was argued that one should include the non-gluon piece inside the hard part of the charge \cite{He:2019pll}.  As we will explicitly discuss later, we start from the linear charge \eqref{eq:Q_lin} acting on asymptotic states however such a hard contribution in our case is indeed present as it will be generated by the time evolution given by the M\o ller operator. At $\mathcal{I}^-$ there is an analogous construction where the linearised asymptotic charge is given by
\begin{align}
Q^{\text{lin}}_{\epsilon} = \int_{\mathcal{I}^{-}} d^2 z d v ~ \epsilon^a(z, \bar{z}) \left[ \partial_v (\partial_z A^a_{\bar{z}} + \partial_{\bar{z}} A^a_z) \right]
\end{align}
in advanced Bondi coordinates.

We will be interested in understanding the action of these charges on scattering states and in order to hew closely to the standard S-matrix formulation we will start with the usual equal-time commutators on a space-like hypersurface and by taking the $t \to \pm \infty$ limit of the usual free field expansion
\begin{align}
\mathcal A_{\mu}^{a}(x) &=  \sum_{\sigma = \pm} \int \widetilde{dq}  \left[\bar{\varepsilon}_{\mu}^{\sigma}(\textbf{q}) a^a_{\sigma}(\textbf{q}) e^{iq\cdot x} + \varepsilon_{\mu}^{\sigma}(\textbf{q}) a^{\dagger a}_{\sigma}(\textbf{q}) e^{-iq\cdot x}\right]~.
\end{align}
The sum is over helicities with polarisation vectors $\varepsilon^\sigma_\mu$ and the equal time commutation relations are given in Appendix \ref{sec:appendixA0}.
We can then evaluate the fields arbitrarily close to $\mathcal{I}^+$ by using the saddle point approximation in the $r\to \infty$ limit \cite{He:2014cra, Strominger:2017zoo,book:Miller,book:deBrujin}  such that, 
\begin{align}
A_z^{a} &= \frac{-i}{8 \pi^2} \frac{\sqrt{2} }{(1 + z \bar{z})} \int_0^{+\infty} d \omega_q  \left[a^a_{+}(\omega_q \hat{x}) e^{-i \omega_q u} - a^{a, \dagger}_{-}(\omega_q \hat{x}) e^{i \omega_q u} \right] \nonumber \\
A_{\bar{z}}^{a} &= \frac{-i}{8 \pi^2} \frac{\sqrt{2} }{(1 + z \bar{z})} \int_0^{+\infty} d \omega_q \left[a^a_{-}(\omega_q \hat{x}) e^{-i \omega_q u} - a^{a, \dagger}_{+}(\omega_q \hat{x}) e^{i \omega_q u} \right]~,
\label{eqn:A_expansion}
\end{align}
where $q^0=\omega_q$ and $\hat{x}=\vec{x}/r$. 
The asymptotic charges can now be expressed in terms of the creation and annihilation operators and for the linear, or soft, charge we have
\<
\hspace{-10pt}Q^{\text{lin}}_\epsilon &=& \lim_{\omega_q \to 0}   \frac{ \omega_q}{4 \pi } \int d^2 z \sqrt{\gamma_{z \bar{z}}} \Bigg[\partial_z \epsilon^a(z, \bar{z}) \left(a^a_{-}(\omega_q \hat{x}) + a^{a\dagger}_{+}(\omega_q \hat{x})\right) 
+\text{h.c.} \Bigg]~ \nonumber \\
\label{eqn:linear_charge}
\>
which provides the connection with the action discussed in the introduction eq.\eqref{eqn:Qcharge_lin}.
The non-linear part of the charge is more complicated to interpret in the quantum theory as it requires normal ordering. For the gluon part we have
\begin{align}
Q^{\text{non-lin}}_\epsilon \Big|_{\text{gluon}}=&  ~i\cg f^{a b c}  \int_{\mathcal{I}^{+}} \frac{d^2 z  }{2(2\pi)^3} \gamma_{z \bar z}  \epsilon^a(z, \bar{z})  \Big\{\int_0^{+\infty} d \omega_q~ \omega_q \sum_{\sigma=\pm} a^{b\dagger}_{\sigma }(\omega_q \hat{x}) a^{c}_{\sigma }(\omega_q \hat{x}) \nn \nonumber \\
&\kern-45pt  +  \frac{1}{2} \int_0^{+\infty} \prod_{i = 1}^2 d \omega_{q_i} \delta( \omega_{q_1} + \omega_{q_2}) ( \omega_{q_1}-\omega_{q_2})  \left[ a^{b}_{+}(\omega_{q_1} \hat{x}) a^{c}_{-}(\omega_{q_2} \hat{x}) - a^{b\dagger}_{-}(\omega_{q_1} \hat{x}) a^{c\dagger}_{+}(\omega_{q_2} \hat{x}) \right]\Big\}\nn
\label{eqn:nonlinear_charge}
\end{align}
where we have normal ordered the expression and dropped the resulting constant.  As the energy integral is over only positive values, the delta-function in the second term has no non-zero support in the integration region and it is therefore vanishing. Defining the gluon number density operator 
\<
\rho^a_g(\vec{q})=-i f^{abc} \sum_{\sigma=\pm} a^{b\dagger}_\sigma(\omega_q \hat{x}) a^{c}_\sigma(\omega_q \hat{x})
\>
we can write the first term in the particularly simple form
\<
Q^{\text{non-lin}}_\epsilon \Big|_{ a^\dagger a-\text{gluon}}=&-\cg\int \widetilde{dq}~ \rho_g^a(\vec{q}) \epsilon^a(\hat{x})~.
\label{eq:Q_nonlin_osc}
\>

There is one important subtlety in relating the linear charge to large gauge transformations. Using the equal-time commutators and the expression \eqref{eqn:linear_charge} one can show that
\<
\label{eq:charge_comm}
& & [Q_\epsilon^{\text{lin}}, a^a_+(\vec{q})] =-\tfrac{(2\pi)^2\delta(\omega_q)}{\sqrt{\gamma_{z\bar{z}} }} \partial_z \epsilon^a(z, \bar z)\\\
& &[Q_\epsilon^{\text{lin}}, a^{a \dagger}_+(\vec{q})] =\tfrac{(2\pi)^2\delta(\omega_q)}{ \sqrt{\gamma_{z\bar{z}} }} \partial_{\bar z} \epsilon^a(z, \bar z) \\
& & [Q_\epsilon^{\text{lin}}, a^a_-(\vec{q})] =-\tfrac{(2\pi)^2\delta(\omega_q)}{ \sqrt{\gamma_{z\bar{z}} }} \partial_{\bar z} \epsilon^a(z, \bar z) \\
& & [Q_\epsilon^{\text{lin}}, a^{a \dagger}_-(\vec{q})] =\tfrac{(2\pi)^2\delta(\omega_q)}{ \sqrt{\gamma_{z\bar{z}} }} \partial_{ z} \epsilon^a(z, \bar z)~.
\>
Thus the gauge transformation of the asymptotic field  $A_z^{a}(u, z, \bar z)$ is given by
\<
& & [Q^{\text {lin}}_\epsilon, A_z^{a}(u, z, \bar z)]=\tfrac{i}{2} \partial_z \epsilon^a( z, \bar z))  \int_0^\infty d\omega_q ~\delta(\omega_q)(e^{-i \omega_q u}+e^{+i \omega_q u})~.
\>
This will  give the incorrect large gauge transformations using the usual definition of the delta-function but it can be remedied by inserting a factor of two for the zero-mode contribution. This is related to the fact noted in the context of the Abelian theory \cite{He:2014cra, He:2014laa, Gabai:2016kuf} that
the radiative phase space at $\mathcal{I}^+$ defined as $\Gamma_+ = \{F^a_{u z}, F^a_{u \bar z}\}$ will not give the usual linearised large gauge transformation. It is easy to check that
\begin{align}
[Q^{\text{lin}}_{\epsilon}, A_z^a(u,z,\bar z)] = \frac{i}{2} \partial_z \epsilon^a(z, \bar z) \neq i \delta_{\epsilon} A_z^a(u,z, \bar z)
\end{align}
using the standard Poisson brackets of the non-Abelian gauge theory
\begin{align}
[A_z^a(u,z,\bar z), A_w^b(u^\prime,w,\bar w)] = - \frac{i }{4} \delta^{a b} \Theta(u - u^{\prime}) \delta^2(w -z)~.
\label{eqn:quantization}
\end{align}
The problem is that the zero modes at the boundary correspond to a single real scalar field and not to a complex one: in the zero frequency limit the two helicities are identified and there is a miscounting of the degrees of freedom. This problem can be solved as in \cite{He:2014cra, He:2014laa} by imposing additional constraints at the boundaries of $\mathcal{I}^+$.  Alternatively one can add the factor of two for zero modes as was done by the authors of \cite{Gabai:2016kuf} used in the definition of the QED coherent state operator. We will follow a similar procedure however in our case we modify the charge, adding a factor of two for zero-modes, and leave the coherent state, and the amplitudes, unchanged. 

\paragraph{Splitting of the charge contributions according to the energy scale E}

A key point is that once we pick a scale $E$ we split the Fock space of free particles into the hard and the soft parts $\mathcal{H}= \mathcal{H}^E_s \otimes \mathcal{H}^E_h$. Indeed we could make a splitting of the energy integral to separate the two contributions
\begin{align}
\int_\lambda^{+\infty} d \omega_q \to \int_{\lambda}^E d \omega_q + \int_{E}^{+\infty} d \omega_q ~, 
\end{align}
where we have introduced $\lambda$ as an infrared cutoff.
As mentioned previously, we will in fact take the linearised charge in the asymptotic region as our starting point and see how the non-linearity emerges from commuting with the  evolution operators. As we will see this corresponds to effectively having
\begin{align}
Q^{\text{lin}}_{\epsilon} &= Q^{\text{lin}}_{\epsilon,s} \otimes \mathbb{I}_h  ~~~\text{and}~~~
Q^{\text{non-lin}}_{\epsilon} = \mathbb{I}_s \otimes Q^{\text{non-lin}}_{\epsilon,h}\Big|_{\text{gluon}+\text{quarks}} ~,
\end{align}
where we have a non-linear hard term for both gluons and quarks. This is similar to the approach taken in \cite{He:2019pll}.


\section{Asymptotic Hamiltonian and Soft Evolution Operators}

The starting point for the coherent state approach to IR divergences \cite{Kulish:1970ut} is the choice of an appropriate asymptotic Hamiltonian describing the parton dynamics in the far future and far past. We will review,  following the arguments of \cite{CIAFALONI1985}, how one can carry out the Faddeev-Kulish approach in the non-Abelian case at the leading order in the IR divergences. We start from the splitting of the standard QCD interaction Hamiltonian in the interaction representation into soft and hard parts:
\begin{align}
H^I (t) = H_h^{E}(t) + H_s^{E}(t)~.
\end{align}
This is done by introducing  at each interaction vertex the energy transfer $\nu$ 
\begin{align}
\nu = \left| \sum_{i} \eta_i \omega_i \right| \qquad \sum_{i} \eta_i \textbf{q}_i = 0~,
\label{eq:cutoff}
\end{align}
where $\omega_i$ are the energies of the interacting particles with $\eta_i = +1$ (resp.\ $-1$) for incoming (resp.\ outgoing) particles. We define the soft part of the Hamiltonian as containing only energies below a scale $\nu< E$ and we also introduce a lower cut-off $\lambda < \nu$. The lower energy bound $\lambda$ is not only required by our use of perturbative QCD but also to have a good definition of the FK states \cite{Kulish:1970ut} however in the following we will sometimes leave it implicit. It is important to note that the region $\nu<E$ contains both soft and collinear subregions. For example \cite{DELDUCA1989} considered a cubic vertex with incoming gluon energy $\omega_1=|\mathbf{q}_1|$ and outgoing energies $\omega_i=|\mathbf{q}_i|$ for $i=2,3$. In particular $\omega_3=\sqrt{\omega_1^2+\omega_2^2 -2 \omega_1\omega_2 \cos\theta_{12}}$, where $\theta_{12}$ is the angle between $\mathbf{q}_1$ and $\mathbf{q}_2$. The condition \eqref{eq:cutoff} then defines a hyperbola in $\omega_2$--$\cos \theta_{12}$ plane 
\begin{align}
\cos \theta_{12} =-\left(1+\frac{E}{2\omega_1}\right)\frac{E}{\omega_2}+\left(1+\frac{E}{\omega_1 }\right)
	\end{align}
which defines regions of small $\omega_2$ and $\cos \theta_{12}\sim 1$. Alternatively one can define the asymptotic region by imposing both an angular cutoff $\theta_{12}<\Theta$ and an energy cutoff $\omega_2<M$. In this way either form of the cutoffs can be used to treat the soft and collinear regions for any of the particles. 

The soft M\o ller operators are defined as the standard time-ordered product
\begin{align}
\Omega^E_{\pm} = T \text{exp}\left[-i \int_{\mp \infty}^0 H_s^{E}(t) d t \right]~,
\end{align}
but in order to isolate the leading IR singularities, which are of the form $(\cg^{2} \log \left(\frac{E}{\lambda} \right))^k$ at $k$-loops, it is useful to transform to frequency space and so write the Hamiltonian as \cite{CATANI1986}
\begin{align}
H_s^{E}(t) = \sum_{\eta = \pm} \int_{\lambda}^E d \nu~ h^{\eta}(\nu) e^{-i \eta \nu t}~.
\label{eqn:soft}
\end{align}
Using this expression we can write the soft-evolution operator as 
\begin{align}
\Omega^E_{\pm} =\sum_{n=1}^\infty & \sum_{\eta_i=\pm}\int_\lambda^E d\nu_1 \dots d\nu_n \frac{h^{\eta_n}(\nu_n)\dots h^{\eta_1}(\nu_1)}{(\eta_n \nu_n+\dots +\eta_1 \nu_1\pm i 0 )\dots (\eta_1 \nu_1 \pm i 0)}~.
\end{align}
Infrared singularities come from vanishing energy denominators, and in particular leading logarithms come from the region specified by the strong ordering
\begin{align}
\lambda \ll \nu_1 \ll \nu_2 \ll \dots \nu_n \ll E~,
\end{align}
where we can thus approximate
\begin{align}
&\frac{1}{\eta_n \nu_n + \dots + \eta_1 \nu_1 \pm i 0} \cdot .... \cdot \frac{1}{\eta_1 \nu_1 \pm i 0} \simeq \prod_{i = 1}^n \frac{1}{\eta_i \nu_i} \Theta(\nu_n > \dots  > \nu_1)~.
\label{eqn:ordering}
\end{align}
To leading order (double pole accuracy) in the IR divergences, for which we can neglect the distinction between $\Omega^E_+$ and $\Omega^E_-$ due to the $i 0$ prescription at leading order, the M\o ller operator is then given by the frequency-ordered exponential
\begin{align}
\Omega^E &= \mathcal{P}_{\nu} \text{exp}\left(\int_{\lambda}^E  \sum_{\eta = \pm} \frac{d \nu}{\eta \, \nu} h^{\eta}(\nu)\right)~.
\end{align}
Furthermore, it is possible to show that at this order in the IR:
\begin{itemize}
	\item  In each three-gluon vertex we may assume there is always one gluon which is much softer than the others and so we can use the eikonal approximation,
	\item Quarks interact only via eikonal vertices and pair production is neglected because the process $g \rightarrow q \bar{q}$ is IR finite,
	\item Four-gluon vertices and ghost contributions can be neglected.
\end{itemize}
With these assumptions it is possible to write the soft interaction Hamiltonian as a sum of two terms, one depending on quarks and a soft gluon and a second purely gluonic cubic term
\begin{align}
H_s^{E}(t) = H_{ffg}^{E}(t) + H_{ggg}^{E}(t)~.
\end{align}
These can be combined so that
\begin{align}
&H_s^E (t) = -g_{\text{\tiny YM}} \int_{\omega_q} \widetilde{d p} \int^E_\lambda \widetilde{d q}~ \rho^a (\textbf{p}) \hat{p} \cdot [a^a(\textbf{q}) e^{i \hat{p} \cdot q t} + h.c.] \nn\\
&~~~\text{with} ~~~\rho^a(\textbf{p}) =\rho_f^a(\textbf{p})+\sum_\sigma a^{\dagger b}_\sigma(\textbf{p}) T_A{}^a{}_{bc} a^{c}_\sigma(\textbf{p})~,
\label{eqn:H_final}
\end{align}
where $\rho$ now contains a contribution not only from fermionic matter $\rho_f$ but also from the hard gluons with energies $\omega_p> \omega_q$.  It is important to note that while the density operator involves gluons which are harder than the soft gluon at that vertex it does not commute with all soft gluon operators. This is fundamentally different from the QED case and corresponds to the non-linear nature of the gauge symmetry.
\begin{figure}
	\centering
	$
	\includegraphicsbox[scale=1.0]{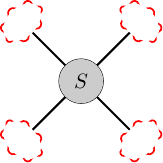}~~
	=~~\includegraphicsbox[scale=1.0]{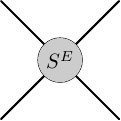}$
	\caption{The dressing of each external leg, represented here by the red cloud, removes the IR singularities and produces an IR finite S-matrix. }
	\label{fig:SmatrixRel}
\end{figure}
Using the Hamiltonian $H_s^E(t)$ in \eqref{eqn:H_final} the soft M\o ller operator becomes
\<
\label{eq:softev}
\Omega^E
&=&\mathcal{P}_\omega \text{exp}\Big[ \int_\lambda^E \widetilde{dq}~ \mathcal{J}_{q} \cdot \Pi_q\Big]~,
\>
where  $\Pi^a_\mu(q)=a_\mu^a(q)-a^{a \dagger}_\mu(q)$ is the displacement operator and
\<
\mathcal{J}^a_{q}{}_\mu= \cg  \int_{\omega_q} \widetilde{dp}~ \rho^a(\textbf{p})\frac{p_\mu}{p\cdot q}
\>
and the dot product denotes the contraction of both Lorentz and gauge indices where appropriate. The exponential is interpreted as being ordered in the soft gluon energies with smaller energies to the right. Here we have switched from the frequency ordering to an energy ordering. For a complete equivalence to the frequency ordering we should also include an ordering on angles and a cutoff on the angular region. However as we are only interested in the leading IR region - which is both soft and collinear- it is sufficient to only consider the energy cutoff. Where this leads to expressions appearing singular an angular cutoff can be mentally added. For our explicit computations in subsequent sections we in fact use dimensional regularisation which simultaneous treats both soft and collinear singularities but the frequency/energy ordering is useful for the physical picture it provides. 

Finally, the soft evolution operator can be used to define an IR-finite S-matrix $S^E$ by removing the IR singularities due to initial and final state interactions
\<
S^E=\Omega^E_- \,S\, \Omega^{E\dagger}_+
\>
as shown schematically in \figref{fig:SmatrixRel}.


\section{Large Gauge Transformations for Soft Evolution Operators}
\label{sec:LGTSEv}
We are interested in understanding the transformation properties of the S-matrix under large gauge transformations and the corresponding Ward identity for amplitudes. As the S-matrix relates states in the far future and far past, where the theory is taken to be free, we will use the linearised charge, \eqref{eq:Q_lin}, when we compute matrix elements of the commutator 
\begin{align}
[Q_\epsilon^{\text{lin}}, S] := Q_\epsilon^{+,\text{lin}} S - S Q_\epsilon^{-,\text{lin}}~,
\end{align}
where we use the appropriate linearised charge $Q^{\pm}$ for incoming and outgoing states. The key to our approach is to use the soft-evolution operator to relate the free theory to the interacting theory and so relate the linearised charge to the non-linear contributions. As the soft-evolution operator can be defined in the quantum theory, this gives a method of defining the correct quantum non-linear corrections to the charge. This can be done by analysing the gauge transformation properties of the soft-evolution operator. 

\begin{figure}
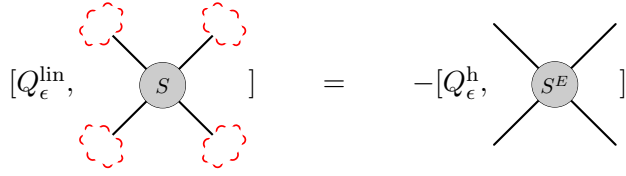
\centering
	$ 
	[Q_\epsilon^{\text{lin}}, \includegraphicsbox[scale=1 ]{}]\qquad=\qquad	-[Q_\epsilon^{\text{h}}, \includegraphicsbox[scale=1]{}]$
	\caption{The Ward identity for the asymptotic charge. The red clouds represent the parton dressing factors comprising soft gluons.}
	\label{fig:TreeQ}
\end{figure}

Before carrying out the computation, we can see how such non-linear terms imply a Ward identity for the S-matrix: If we had that the evolution-operator transformation was given by a relation of the form
\<
\label{eq:Qh}
[Q_\epsilon^{\text{lin}}, \Omega^E]=Q_\epsilon^{\text{h}}\Omega^E
\>
for some non-linear charge $Q_\epsilon^{\text{h}}$, then we would have the relation
\begin{align}
&\langle\!\langle\{p_f, \alpha_f\}\|\,[Q_\epsilon^{\text{lin}}, S]\,\|\{p_i, \alpha_i\}\rangle\!\rangle=\langle\{p_f, \alpha_f\}|[Q_\epsilon^{\text{lin}} - Q_\epsilon^{\text{h}}, S^E]|\{p_i, \alpha_i\}\rangle~.
\end{align}
Here we have used dressed states, e.g.
\<
\|\{p_i, \alpha_i\}\rangle\!\rangle=\Omega^{E\dagger}\prod_{i\in\text{in}} b^\dagger_{\alpha_i}(p_i)|0\rangle~,
\>
in computing matrix elements. 
If the linearised charge produces states orthogonal to scattering states constructed on the original vacuum, this becomes 
\begin{align}
 & \langle\!\langle\{p_f, \alpha_f\}\|\,[Q_\epsilon^{\text{lin}}, S]\,\|\{p_i, \alpha_i\}\rangle\!\rangle=-
\langle\{p_f, \alpha_f\}|[Q_\epsilon^{\text{h}},S^E]|\{p_i, \alpha_i\}\rangle~. 
\end{align}
This relation, graphically represented in \figref{fig:TreeQ}, is what we refer to as the Ward identity. 

Our goal is to study this relation in the context of QCD using the soft-evolution operators and capturing the leading IR singularities. To this end we consider the explicit expression for the linearised large gauge transformations using the definition of the linearised charge \eqref{eq:Q_lin} with the  commutators \eqref{eq:charge_comm}.
It is convenient to first integrate by parts for the $z$ and $\bar z$ variables and start with the expression 
\<
Q_\epsilon^{\text{lin}}=- \int_{\mathcal{I}^+} d^2z du \big[\partial_z \epsilon^a \partial_u  A^a_{\bar z}+\partial_{\bar z} \epsilon^a \partial_u  A^a_z \big]~.
\>
 In principle we must carefully account for any boundary terms that occur in this step, however at the end of the computation  we will undo this integration by parts and so remove the boundary terms again. 
Using the definition of the energy ordering, we  calculate
\begin{align}
&[Q_\epsilon^{\text{lin}}, \Omega^E]=\sum_n \Big[Q_\epsilon^{\text{lin}}, \int_{\omega_{n-1}} \widetilde{dq}_{n} \dots \int_\lambda^E \widetilde{dq}_1~ \mathcal{J}_{n} \cdot \Pi_n\dots \mathcal{J}_{1} \cdot \Pi_1\Big]~.
\end{align}
As the charge only involves zero modes, it acts only on the last term in each element of the sum, i.e.\ $\int_\lambda^E \widetilde{dq}_1 \mathcal{J}_{1} \cdot \Pi_1$, and then only for $\omega_1=\lambda\sim 0$. That is,  we are assuming that the charge acts on the mode with energy $\omega_{1}=\lambda\sim 0$  but not on $\omega_{2}> \omega_{1}$. By straightforward computation one finds
\<
[Q_\epsilon^{\text{lin}}, \int_\lambda^E \widetilde{dq} ~\mathcal{J}\cdot \Pi]
&=&\mathcal{N}(\epsilon)~,
\>
where $\mathcal{N}(\epsilon)=\int_\lambda \widetilde{dp}~ \rho^a(\textbf{p}) N^a(\epsilon, \textbf{p})$
with
\begin{align}
&N^a(\epsilon,\textbf{p})=-(2\pi)^2 \cg \int \frac{\widetilde{dq}}{\sqrt{\gamma_{z \bar{z}}}}\delta(\omega_q)
\frac{2p}{p\cdot q}\cdot(\varepsilon^-(q)\partial_z \epsilon^a(q)+\varepsilon^+(q) \partial_{\bar z} \epsilon^a(q))~.  
\end{align}
This expression for $N^a$ depends on having performed an integration by parts and so ignores any potential boundary terms. For example it can be seen that such a term vanishes for the case $\epsilon^a$ being a constant. One can undo the integration by parts, and so recover the dropped boundary terms, 
and write 
\begin{align}
N^a(\epsilon,\textbf{p})=\frac{\cg}{2\pi }\int d\omega_q \delta(\omega_q) \int d^2z \gamma_{z\bar z}~\frac{-p^2}{(p\cdot \hat{q})^2}\epsilon^a(z,\bar z)~,  
 \label{eq:Nep}
\end{align}
where we use the notation $\hat{q}^\mu=q^\mu/\omega_q$ and which for the case of constant $\epsilon$ becomes $N^a(\epsilon,\textbf{p})= \cg \epsilon^a \int d\omega \delta(\omega)$. 
In these expressions we must interpret the $\delta$-functions as taking the soft limit in the appropriate fashion. 
We have thus found that 
\<
[Q_\epsilon^{\text{lin}}, \Omega^E]
&=& \Omega^E \mathcal{N}(\epsilon)~,
\>
which is almost what we would have expected from the classical analysis of the non-linear charge. However as
\<
[\mathcal{J}_j \cdot \Pi_j, \int_{\lambda} \widetilde{dp} ~\rho^b(\textbf{p}) N^b(\epsilon, \textbf{p}) ]=0
\>
only if $N^b(\epsilon, \textbf{p})=N^b(\epsilon, \textbf{q}_j)$, we must correctly order the terms to put the expression in the form \eqref{eq:Qh}. Performing the non-trivial commutators to one-loop, $\mathcal{O}({\cg^2})$, we find 
\<
[Q_\epsilon^{\text{lin}}, \Omega^E]= \big[\mathcal{N}(\epsilon)+\mathcal{A}(\epsilon)\big]\Omega^E~,
\>
where
\begin{align}
&\mathcal{A}(\epsilon)=
i g_{\text{\tiny YM}} f^{abc}\int_\lambda^E \widetilde{dq}\int_\omega \widetilde{dp}~ \rho^a(\textbf{p})
\frac{p\cdot \Pi^c(q)}{p\cdot q} \cdot \big[N^b(\epsilon, \textbf{p})-
N^b(\epsilon, \textbf{q})\big]~.
\end{align}
Here we see that there are corrections to the tree-level expression for $Q_\epsilon^{\text{h}}$ which arise due to the non-Abelian nature of the theory. Indeed we can see that there will be further non-linear corrections at each higher loop orders. These corrections do vanish in the case of a constant $\epsilon$, which is to say that the total colour charge does not receive any corrections. Moreover the correction involves a soft-gluon operator which we might expect to have vanishing contribution when computing matrix elements of the IR-finite S-matrix. We now turn to the computation of exactly such matrix elements.

\section{Ward Identities for Dressed S-matrix}
To  compute matrix elements between dressed states, we make use of the non-trivial fact that, to leading order in the soft divergence, the dressing of external states factorises in colour space  \cite{CATANI1985, CIAFALONI1985, CATANI1986}
\begin{align}
\|\{p_i,\alpha_i\}\rangle\!\rangle\equiv\Omega^{E\dagger} |\{p_i,\alpha_i\}\rangle=\prod_{i\in \text{in}} \mathcal{U}^{p_iE}_{\alpha_i\beta_i}(\Pi) b^\dagger_{\beta_i}(p_i) |0\rangle~,
\end{align}
where the coherent-state operator $\mathcal{U}^{p_iE}_{\alpha_i\beta_i}(\Pi)$ is a functional of the soft gluons only. The coherent-state operator for a parton in the gauge group representation with generators $t_{\alpha\beta}^a$ is defined by the energy-ordered integral
\<
\label{eq:CoherentOP}
\mathcal{U}_{\alpha\beta}^{pE}=P_\omega \text{exp}\Big[-g_{\text{\tiny YM}}\int^E_\lambda \widetilde{dq}~\frac{p\cdot \Pi^a_\omega(q)}{p\cdot q} t^a\Big]_{\alpha\beta}~,
\>
where the dressed gluon field is similarly defined by 
\<
\Pi^a_\omega(q)=\mathcal{U}_{ab}^{qE}\Pi^b(q)
\>
and $\mathcal{U}_{ab}^{pE}$ is the adjoint coherent-state operator.  
These non-linear equations can be solved iteratively so that to $\mathcal{O}(\cg^2)$ we have
\begin{align}
\mathcal{U}^{pE}_{\alpha\beta}=&~\delta_{\alpha\beta} - 
\cg \int_\lambda \widetilde{dq}~ \frac{p\cdot \Pi^e(q)}{p\cdot q} t^e_{\alpha \beta} 
\nn\\
&+\cg^2 \int_{\lambda} \widetilde{dq}_1\int_{\omega_1} \widetilde{dq}_2
\left(
\frac{p\cdot \Pi^{e_2}(q_2)}{p\cdot q_2} t^{e_2}_{\alpha \gamma}
\right)
\left(
\frac{p\cdot \Pi^{e_1}(q_1)}{p\cdot q_1} t^{e_1}_{\gamma \beta}
\right)\nn\\
&-\cg^2  \int_\lambda \widetilde{dq}_1\int_{\omega_1} \widetilde{dq}_2
\left(
\frac{p\cdot \Pi^{e_2}(q_2)}{p\cdot q_2}
\right)
\left(
\frac{q_2\cdot \Pi^{e_1}(q_1)}{q_2\cdot q_1} 
\right) \cdot [t^{e_2},t^{e_1}]_{\alpha \beta}~.
\end{align}
This dressing factor captures the leading-order effects of soft-gluon radiation of each of the hard partons. It includes all-order effects arising from arbitrary numbers of gluons being radiated, as well as loop effects which arise from normal ordering each of the terms.  

\paragraph{Tree-level Ward Identity} While the calculation is essentially identical to the Abelian case, to fix conventions we will start with the tree-level result. The relevant terms for computing matrix elements of the commutator of the linearised charge with the S-matrix are:
\begin{align}
&\langle\!\langle \{p_f,\alpha_f\} \|\,[Q^{\text{lin}}_\epsilon,S] \,\|\{p_i,\alpha_i\}\rangle\!\rangle= \langle 0|\prod_{f\in \text{out}}  b_{\alpha_f}(p_f)\Bigg([Q_\epsilon^{\text{lin}},  S]  \nn \\
&- g_{\text{\tiny YM}} \int_\lambda \widetilde{dq}  \Big[\sum_{\ell\in \text{out}}\frac{p_\ell \cdot \Pi^{e}(q) }{p_\ell\cdot q}~t^{e}_{\ell} ~ [Q_\epsilon^{\text{lin}}, S] + \sum_{\ell\in \text{in}}[Q_\epsilon^{\text{lin}}, S]\frac{p_\ell \cdot \Pi^{e}(q) }{p_\ell\cdot q}t^{e}_{\ell}\Big]\Bigg)\prod_{i\in\text{in}} b^\dagger_{\alpha_i}(p_i) |0\rangle~,
\end{align}
where the subscripts on the colour generators indicate the parton leg upon which they act and for convenience we denote, for example, $t^a_{\alpha_j\beta_j} b^\dagger_{\beta_j} =t^a_jb^\dagger_{\alpha_j}$.
\begin{figure}
	\begin{eqnarray}
	\begin{array}{cc}
	\includegraphicsbox[scale=1]{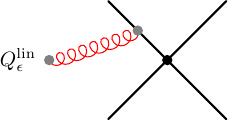}
	& ~~~~~~~	\includegraphicsbox[scale=1]{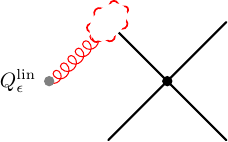}\nn\\
	& \nn\\
	~~~~~~~~~~~~~~~~~(4.\text{a}) &~~~~~~~~~~~~~~~~~~~~~~~~(4.\text{b})
	\end{array}
	\end{eqnarray}
	\caption{The tree-level contributions to the Ward identity.}
	\label{fig:TreeSoft}
\end{figure}

The two contributions to the matrix element can be represented graphically as in \figref{fig:TreeSoft}. The first term, corresponding to diagram $(4.\text{a})$, corresponds to the usual absorption or emission of the soft gluon in the linearised charge. We compute these terms by using the soft-gluon theorem, see \eqref{eq:tree_soft} and \eqref{eq:treesoftcurr},
and due to the sign difference in the soft limits of incoming and outgoing gluons the two terms of the commutator add rather than cancel. 
The next terms, corresponding to diagram $(4.\text{b})$ in \figref{fig:TreeSoft}, arise from contracting oscillators in the linearised charge with those in the coherent states. These terms cancel due to the sign from the commutator. Thus we find 
\begin{align}
&\langle\!\langle \{p_f,\alpha_f\}\|\, [Q_\epsilon^{\text{lin}}, S]\,\|\{p_i,\alpha_i\}\rangle\!\rangle= -\Big[\sum_{\ell\in \text{out} } Q_\epsilon^{\text{h}}(p_\ell)-\sum_{\ell\in \text{in} } Q_\epsilon^{\text{h}}(p_\ell)\Big]~\mathcal{M}^{(0)}_n~, 
\end{align}
where we have introduced the eigenvalue of the non-linear charge $Q_\epsilon^{\text{h}}(p)=Q^{h,a}_\epsilon(p) t^a$ where
\begin{align}
Q^{h,a}_\epsilon(p)
=&-8\pi^2 \cg \int \widetilde{dq}   \frac{\delta(\omega) }{ \sqrt{\gamma_{z\bar z}}}\cdot \Big[
\partial_z \epsilon^a(\hat{q}) \frac{  \varepsilon^-\cdot p }
{q\cdot p}
+\partial_{\bar z} \epsilon^a(\hat{q}) \frac{ \varepsilon^+\cdot p}{q\cdot p}\Big]
\label{eq:Q1_ev}
\end{align}
which, as expected,  is $N^a(\epsilon, \textbf{p})$ as defined in \eqref{eq:Nep}.

As in the QED case the charge acts on the Fock vacuum to produce a state orthogonal to all scattering states built on the original vacuum.
This can be seen by keeping only those terms where the charge corresponds to an emitted gluon, and for convenience considering only incoming hard partons. While we now only have half the terms from the commutator, we can still have contributions of the form represented in \figref{fig:TreeSoft}, however due to a cancellation between diagrams $(4.\text{a})$ and $(4.\text{b})$, one finds
\<
\langle\!\langle0\|\,Q_\epsilon^{\text{lin}}S \, \|\{p_i,\alpha_i\}\rangle\!\rangle&=&0~
\>
at tree-level. 

\paragraph{Finite One-loop S-matrix}Now let us proceed to a one-loop calculation, and again for now considering only in-particles, we compute the one-loop matrix elements
\begin{align}
\langle\!\langle 0\|\, S\,\|\{p_i,\alpha_i\}\rangle\!\rangle &~=\langle 0|S \prod_i \mathcal{U}^{p_iE}_{\alpha_i\beta_i} b^\dagger_{\beta_i}(p_i) |0\rangle\nn\\
&~= \langle 0| S\prod_i  b^\dagger_{\alpha_i}(p_i) |0\rangle\nn\\
&~~- g_{\text{\tiny YM}} \sum_j t^{e_1}_j \int_\lambda \widetilde{dq}  \langle 0| S~ \tfrac{p_j \cdot \Pi^{e_1}(q) }{p_j\cdot q}
\prod_i b^\dagger_{\alpha_i}(p_i) |0\rangle\nn\\
&~~+\frac{g_{\text{\tiny YM}}^2}{2} \sum_{j\neq k}  t^{e_1}_j t^{e_2}_k
\int_\lambda \widetilde{dq}_1\int_\lambda \widetilde{dq}_2 \langle 0| S
\tfrac{p_j\cdot \Pi^{e_1}(q_1)}{p_j\cdot q_1} 
\tfrac{p_k\cdot \Pi^{e_2}(q_2)}{p_k\cdot q_2} \prod_i b^\dagger_{\alpha_i}(p_i)|0\rangle\nn\\
&~~+ \text{one-parton terms~.}
\end{align}

\begin{figure}\centering
	\begin{eqnarray}
	\begin{array}{ccc}
	\left\{\includegraphicsbox[scale=1]{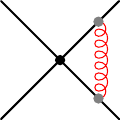}+\dots\right\}&-\includegraphicsbox[scale=0.95]{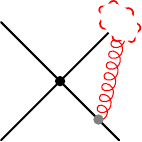}& - \includegraphicsbox[scale=0.95]{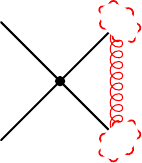}=0\nn\\
	& \nn\\
	\hspace{-5pt}(5.\text{a})~~~~~~ &(5.\text{b})&\hspace{-15pt}(5.\text{c})
	\end{array}
	\end{eqnarray}
	\caption{The IR-divergent contributions at one-loop involving two external partons.}
	\label{fig:IRdvg2pt}
\end{figure}

\begin{figure}\centering
	\begin{eqnarray}
	\begin{array}{ccc}
	\includegraphicsbox[scale=1.0]{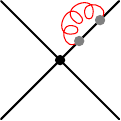}~~~~~~~~~~~~~ & 	\includegraphicsbox[scale=0.95]{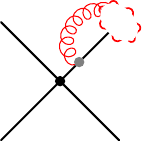} ~~~~~~~~~& 	\includegraphicsbox[scale=0.95]{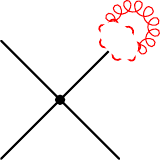}\nn\\
	& & \nn\\
	\hspace{-38pt}(6.\text{a})&\hspace{-32pt}(6.\text{b})&\hspace{-16pt}(6.\text{c}) 
	\end{array}
	\end{eqnarray}
	\caption{The IR divergent contributions at one-loop involving one external parton.}
	\label{fig:IRdvg1pt}
\end{figure}
The first term is the usual IR-divergent S-matrix  element which arises from diagrams such as $(5.\text{a})$ in \figref{fig:IRdvg2pt}, while the remaining terms are the compensating IR-divergent terms from the coherent state which are graphically represented in diagrams $(5.\text{b})$ and $(5.\text{c})$. Here we will focus on those diagrams which involve two external partons and neglect one-parton contributions such as those in \figref{fig:IRdvg1pt}. These one-parton contributions are subleading in the IR divergences and so are not needed, but can in fact be reconstructed by insisting on gauge invariance. The second term $(5.\text{b})$ gives an $\mathcal{O}(g_{\text{\tiny YM}}^2)$ contribution by using the tree-level soft limit of the S-matrix. 
In particular it can be written as 
\begin{align}
I^{(5.\text{b})} = g_{\text{\tiny YM}}^2 \sum_{j\neq k}   t^{e}_j t^{e}_k \int \widetilde{dq}~  \sum_\sigma \frac{ p_j\cdot \varepsilon^\sigma(q) }{p_j\cdot q}  \frac{p_k \cdot \bar{\varepsilon}^\sigma (q)}{p_k\cdot q}  \mathcal{M}^{(0)}_n~,
\end{align}
where we have again dropped the one-parton, i.e.\ $j= k$, terms.
The next term, corresponding to $(5.\text{c})$, involves contracting two gluons from the dressing factors of two partons. We thus need
\begin{align}
\Pi^{e_j}_\mu(q_1)\Pi^{e_k}_\nu(q_2)~=&~(a_\mu^{e_j}(q_1)-a^{e_j\dagger}_\mu(q_1))(a_\nu^{e_k}(q_2)-a^{e_k\dagger}_\nu(q_2))\nn\\
=&~-\widetilde{\delta}(\textbf{q}_1-\textbf{q}_2) \delta^{e_j e_k}\sum_\sigma \bar{\varepsilon}^\sigma_\mu\varepsilon_\nu^\sigma +\text{terms with two oscillators}
\end{align}
so that we find
\begin{align}
I^{(5.\text{c})} = -\frac{g_{\text{\tiny YM}}^2}{2} \sum_{j\neq k} t^{e}_jt^{e}_k \int \widetilde{dq} \sum_\sigma \frac{p_j\cdot \bar{\varepsilon}^\sigma(q) p_k\cdot {\varepsilon}^\sigma(q)}{p_j\cdot q~ p_k\cdot q} \mathcal{M}^{(0)}_n
\end{align}
which is the same as the $(5.\text{b})$ contribution up to an overall factor. 
We can rewrite the product of polarisation vectors using \eqref{eq:pol_rel} which,
if we include the one-parton contributions and impose total colour conservation to remove the $c_\mu$ dependent terms, becomes 
$\sum_\sigma \bar{\varepsilon}^\sigma_\mu {\varepsilon}^\sigma_\nu=\eta_{\mu\nu}$. 
Hence we find the S-matrix elements
\<
\langle\!\langle 0\|\, S\,\|\{p_i,\alpha_i\}\rangle\!\rangle
&=&\Big[1+ \frac{g_{\text{\tiny YM}}^2}{2} \sum_{j, k}  t^{e}_jt^{e}_k ~I_{jk}\Big] \mathcal{M}^{(0)}_n
\>
in terms of the IR-divergent loop integral
\<
\label{eq:loop_int}
I_{jk}=\int \widetilde{dq} \frac{p_j\cdot p_k}{p_j\cdot q~ p_k\cdot q}~.
\>
While we have previously considered an energy cut-off to regularise divergences, in order to compare with known results it is useful to instead use dimensional regularisation. Keeping the leading divergence in the parameter $\hat{\epsilon}=\tfrac{d}{2}-2$, the loop integral is given by 
\<
I_{jk}=-\frac{1}{2(2\pi)^2\hat{\epsilon}^2}~,
\>
which encodes the double-pole singularity due to both soft and collinear divergences at leading logarithmic accuracy. To one-loop order and to our accuracy the amplitude is known to be, \cite{Bierenbaum:2011gg}, 
\<
\label{eq:1loopamp}
\mathcal{M}_n=\Big[1+\frac{\cg^2}{16\pi^2\hat{\epsilon}^2} \sum_{j\neq k}  t^{e}_jt^{e}_k\Big]\mathcal{M}_n^{(0)}
\>
and hence we see that, as expected, the singular parts cancel in the S-matrix elements. This can be repeated for generic outgoing states and again one finds that the one-loop leading singularities cancel 
\<
\langle\!\langle \{p_f,\alpha_f\}\|\, S \,\|\{p_i,\alpha_i\}\rangle\!\rangle |_{\mathcal{O}(\cg^2)}=0+\mathcal{O}\left(1/\hat{\epsilon}\right)
\>
as expected. An analogous approach to removing the singularities would be the Wilson-line dressing which reproduces the same leading IR-divergences \cite{Gardi:2013ita} (see also \cite{FRENKEL1982,BASSETTO1982189,Laenen:2008gt,Gardi2011,Naculich:2011ry,Oxburgh:2012zr}).

\paragraph{One-loop Ward Identity} 
We now want to compute the one-loop correction to the Ward identity using the dressed states.
\begin{figure*}
	\begin{eqnarray}
	\begin{array}{ccc}
	\left\{	\includegraphicsbox[scale=1]{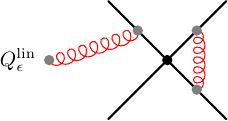} +\dots \right\}&+~~~	\includegraphicsbox[scale=1]{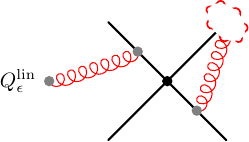}&+~~~	\includegraphicsbox[scale=1]{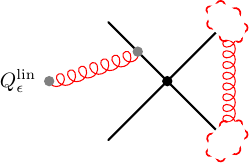}\nn\\
	&  \nn\\
	~~~~~~~~~~(7.\text{a}) &~~~~~~~~~~~~~~~~~~~(7.\text{b}) & ~~~~~~~~~~~~~~~~~~~~ (7.\text{c})
	\end{array}
	\end{eqnarray}
	\caption{The one-loop contributions to the Ward identity. }
	\label{fig:OneLoopSoft}
\end{figure*}
We thus consider the terms
\begin{align}
&\langle\!\langle 0 \|\,[Q_\epsilon^{\text{lin}},S]\, \|\{p_i,\alpha_i\}\rangle\!\rangle |_{\mathcal{O}(g_{\text{\tiny YM}}^3)}=\langle 0| [Q_\epsilon^{\text{lin}},S] \Bigg\{1
- g_{\text{\tiny YM}} \sum_j t^{e_1}_j \int_\lambda \widetilde{dq} ~ \tfrac{p_j \cdot \Pi^{e_1}(q) }{p_j\cdot q}\nn\\
&\hspace{-40pt}\hspace{2cm}+\frac{g_{\text{\tiny YM}}^2}{2} \sum_{j_s}\prod_{s=1}^2 t^{e_s}_{j_s} \int_\lambda  \widetilde{dq}_s 
\tfrac{p_{j_s}\cdot \Pi^{e_s}(q_s)}{p_{j_s}\cdot q_s} +g_{\text{\tiny YM}}^2\sum_{j} [t^{e_1}_j, t^{e_2}_j]
\int_{\lambda} \widetilde{dq}_1\int_{\omega_1} \widetilde{dq}_2~
\tfrac{q_2\cdot \Pi^{e_1}(q_2)}{q_2\cdot q_1} \tfrac{p_j\cdot \Pi^{e_2}(q_2)}{p_j\cdot q_2}
\nn\\
&\hspace{-40pt}\hspace{2cm} -\frac{g_{\text{\tiny YM}}^3}{3!}\sum_{j_s}t^{e_1}_{j_1} t^{e_2}_{j_2}t^{e_3}_{j_3}\prod_{s=1}^3  \int_\lambda  \widetilde{dq}_s 
\tfrac{p_{j_s}\cdot \Pi^{e_s}(q_s)}{p_{j_s}\cdot q_s} 
\nn\\
&\hspace{-40pt}\hspace{2cm} -g_{\text{\tiny YM}}^3\sum_{j_1,j_2}t^{e_1}_{j_1} [t^{e_2}_{j_2}, t^{e_3}_{j_2}]
\prod_{s=1}^2 \int_\lambda \widetilde{dq}_s \tfrac{p_{j_s}\cdot \Pi^{e_s}(q_s)}{p_{j_s}\cdot q_s} \int_{\omega_2} \widetilde{dq}_3
\tfrac{q_2\cdot \Pi^{e_3}(q_3)}{q_2\cdot q_3}
\nn\\
&\hspace{-40pt}\hspace{2cm} -g_{\text{\tiny YM}}^3\sum_{j_1,j_2}t^{e_1}_{j_1} t^{e_3}_{j_2} t^{e_2}_{j_2}
\prod_{s=1}^2 \int_\lambda \widetilde{dq}_s \tfrac{p_{j_s}\cdot \Pi^{e_s}(q_s)}{p_{j_s}\cdot q_s} \int_{\omega_2} \widetilde{dq}_3
\tfrac{p_{j_2}\cdot \Pi^{e_3}(q_3)}{p_{j_2}\cdot q_3} +\text{one-parton terms}\Bigg\}|\{p_i,\alpha_i\}\rangle~,
\end{align}
where as before the subscripts on the generators denote the hard parton upon which they act. This expression simplifies significantly as
the contributions where the gluon in the charge is contracted with a gluon in the dressing of the partons cancel between the two terms in the commutator. Thus we only need to keep the contributions where the charge contracts with the S-matrix as these terms add. The non-vanishing contributions are graphically represented in \figref{fig:OneLoopSoft} where it can be seen that these contributions closely parallel those of \figref{fig:IRdvg2pt}.  The result can be straightforwardly computed using the previous tree-level and one-loop results. For example diagram $(7.\text{b})$ gives
\<
\label{eq:QS1loopb}
I^{(7.\text{b})} = g_{\text{\tiny YM}}^2 \sum_{j, k, r}Q^{h,a}_\epsilon(p_j)~I_{kr} ~t^a_j t^e_k t^e_r \mathcal{M}_n^{(0)}~,
\>
where $Q_\epsilon^{\text{h},a}$ is defined in \eqref{eq:Q1_ev} and $I_{kr}$ is the integral \eqref{eq:loop_int}. The diagram $(7.\text{c})$ gives the same contribution but with a coefficient of $-\tfrac{1}{2}$, so that $I^{(7.\text{c})} = -\frac{1}{2} I^{(7.\text{b})}$. These terms cancel against contributions such as those in diagram $(7.\text{a})$, which can be computed using the soft limit of one-loop amplitudes given in \eqref{eq:1loopsoft}.
As can be seen in, for example, \eqref{eq:QS1loopb} there are contributions with $j \neq k \neq r$ that involve three external hard-parton legs. These contributions straightforwardly cancel against the appropriate terms in the first term of the one-loop soft limit involving the one-loop amplitude, that is the iterated tree term, 
\begin{align}
\sum_j Q^{h,a}_\epsilon(p_j) t^a_j \mathcal{M}^{(1)}_n\simeq \frac{g_{\text{\tiny YM}}^2}{16\pi^2\hat{\epsilon}^2} \sum_{j\neq k\neq r} Q^{h, a}_\epsilon(p_j) t^a_j t^e_k t^e_r \mathcal{M}^{(0)}_n
\end{align}
using \eqref{eq:1loopamp} for the leading singular terms in the amplitude.
\begin{figure}
	\begin{eqnarray}
	\begin{array}{cc}
	~~~~	\includegraphicsbox[scale=1]{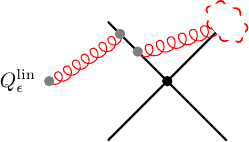}& ~~~~	\includegraphicsbox[scale=1]{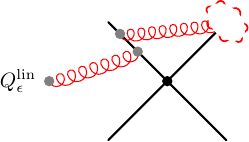}\nn\\
	&  \nn\\
	~~~~~~~~~~~~~~~~~~(8.\text{b.1}) &~~~~~~~~~~~~~~~~~~(8.\text{b.2}) 
	\end{array}
	\end{eqnarray}
	\caption{Ambiguous two-parton contributions to one-loop Ward identity. }
	\label{fig:OneLoopDSoft}
\end{figure}
Slightly more subtle are the contributions which involve only two external legs. 
In particular for the diagram $(7.\text{b})$ there is an ambiguity regarding the order in which one takes the soft limits corresponding to the gluon in the charge and the lowest-energy gluon in the parton dressing. The two orderings are shown schematically in \figref{fig:OneLoopDSoft}. Diagram $(8.\text{b.1})$ corresponds to first taking the soft-limit for the charge before computing the loop integral involving the soft gluon in the dressing factor.  In diagram $(8.\text{b.2})$ we instead take the soft limit for the dressing-factor gluon and then for the gluon in the charge. In particular this gives an ordering of the colour generators 
\<
I^{(8.\text{b}.2)} = - \cg^2 \sum_{j\neq k }Q^{h,a}_\epsilon(p_j)~I_{kr} ~t^e_j t^a_j  t^e_k \mathcal{M}_n^{(0)}
\>
which combines with the other two-parton contributions from $(9.\text{c})$ to cancel against the two-parton contributions from $(7.\text{a})$ which come from the soft limit of the one-loop amplitude.
\begin{figure}
	\begin{eqnarray}
	\begin{array}{cc}
	\includegraphicsbox[scale=1]{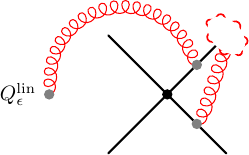} & ~~~~	\includegraphicsbox[scale=1]{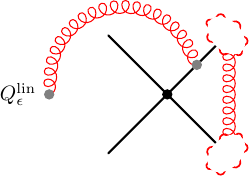}\nn\\
	&  \nn\\
	~~~~~~~~~~~~~(9.\text{b}.3) &~~~~~~~~~~~~~~~~~(9.\text{c}) 
	\end{array}
	\end{eqnarray}
	\caption{Non-ambiguous two-parton contributions to one-loop Ward identity.  }
	\label{fig:OneLoopSoft2part}
\end{figure}
The final result is that using the ordering $(\text{8.b.2})$  we find that the commutator vanishes at one-loop to leading singularity
\<
\langle\!\langle 0 \|\,[Q_\epsilon^{\text{lin}},S] \,\|\{p_i,\alpha_i\}\rangle\!\rangle |_{\mathcal{O}(g_{\text{\tiny YM}}^3)}&=&0+\mathcal{O}(\tfrac{1}{\hat \epsilon})~.
\>
There are in principle contributions which only involve a single external parton momentum which we have not included as they are subleading in the IR expansion, however these can again be found by using gauge invariance. Thus we find that, using this prescription, there are no corrections to the Ward identity. 

If we had chosen the alternative prescription $(8.\text{b}.1)$ we would find a correction which is related to the one-loop soft current and which can be written as 
\begin{align}
&\langle\!\langle 0 \|\,[Q_\epsilon^{\text{lin}},S] \,\|\{p_i,\alpha_i\}\rangle\!\rangle |_{\mathcal{O}(g_{\text{\tiny YM}}^3)}=-
\frac{\cg^2 C_A}{16\pi^2\hat{\epsilon}^2}\sum_{\ell\in \text{in}} Q_\epsilon^{\text{h}}(p_\ell) \mathcal{M}^{(0)}_n~.
\end{align}
In this case we see that there appears to be a one-loop effect which in principle could be interpreted as a correction to the hard charge. Related to this we must also establish that the linearised soft charge acting on the vacuum produces states that are orthogonal to the dressed scattering states at one-loop, that is we must compute  
\<
\langle\!\langle 0 \|\,Q_\epsilon^{\text{lin}}S\,\|\{p_i,\alpha_i\}\rangle\!\rangle |_{\mathcal{O}(g_{\text{\tiny YM}}^3)}~.
\>
\begin{figure*}
	\begin{eqnarray}
	\begin{array}{ccc}
	\left\{	\includegraphicsbox[scale=1]{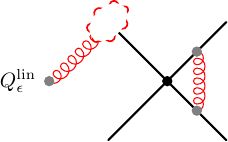}+\dots \right\}&+~~~
	\includegraphicsbox[scale=1]{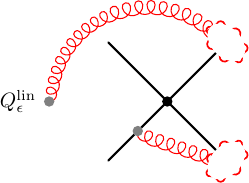}~~~+
	&  \includegraphicsbox[scale=1]{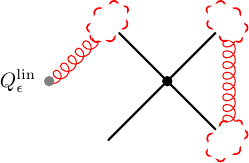}
	\nn\\
	~~~~~~~~~~(10.\text{a})&~~~~~~~~~~~~~~(10.\text{b})&~~~~~~~~~~~~~~(10.\text{c})
	\end{array}
	\end{eqnarray}
	\caption{The additional three parton one-loop contributions to the orthogonality condition. }
	\label{fig:OneLoopOrtho3part}
\end{figure*}
In order to do this, one needs to compute all $\mathcal{O}(g_{\text{\tiny YM}}^3)$ contributions, a calculation which closely parallels the computation of the one-loop soft current \cite{Catani:2000pi}.  There are additional three-parton contributions, see \figref{fig:OneLoopOrtho3part}, which combine, using essentially the manipulations from the tree-level calculation, to zero. That is the contributions from \figref{fig:OneLoopOrtho3part} cancel those of \figref{fig:OneLoopSoft}. 
\begin{figure*}
	\begin{eqnarray}
	\begin{array}{ccc}
	\includegraphicsbox[scale=1]{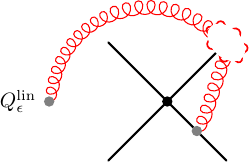} & ~~~~	\includegraphicsbox[scale=1]{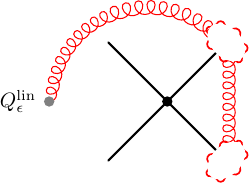}& ~~~\includegraphicsbox[scale=1]{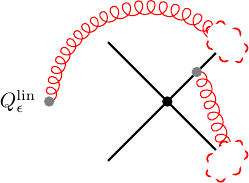}\nn\\
	&  \nn\\
	~~~~~~~~~~~~~~(11.\text{a}) &~~~~~~~~~~~~~~~~~~(11.\text{b}) &~~~~~~~~~~~~~~~~~(11.\text{c})
	\end{array}
	\end{eqnarray}
	\caption{The additional two parton one-loop contributions to the orthogonality condition. }
	\label{fig:OneLoopOrtho}
\end{figure*}
The more non-trivial calculation involves the two-parton contributions (see \figref{fig:OneLoopSoft2part} and \figref{fig:OneLoopOrtho}) and there are contributions with the same ordering ambiguity as in \figref{fig:OneLoopDSoft}. Once again if we make the choice corresponding to $(8.\text{b}.2)$  we find that there is a cancellation of the two-parton contributions so that 
\<
\langle\!\langle 0 \|\,Q_\epsilon^{\text{lin}}S\,\|\{p_i,\alpha_i\}\rangle\!\rangle |_{\mathcal{O}(g_{\text{\tiny YM}}^3)}=0
\>
to leading order in the IR divergences. Alternatively if we choose the ordering $(8.\text{b}.1)$ of \figref{fig:OneLoopDSoft}, we find that all the diagrams that arise from dressing the tree-level amplitude cancel amongst themselves. The diagram corresponding to dressing the one-loop amplitude, $(10.\text{a})$ of \figref{fig:OneLoopOrtho3part}, cancels the contribution due to the  iterated tree term in the soft-limit of the loop amplitude \eqref{eq:1loopsoft}, so that the term arising from the one-loop soft current remains. Thus we find
\begin{align}
\langle\!\langle 0 \|\,Q_\epsilon^{\text{lin}}S\,\|\{p_i,\alpha_i\}\rangle\!\rangle |_{\mathcal{O}(g_{\text{\tiny YM}}^3)}=-
\frac{\cg^2 C_A}{16\pi^2\hat{\epsilon}^2}\sum_{\ell\in\text{in}} Q_\epsilon^{\text{h}}(p_\ell) \mathcal{M}^{(0)}_n~,
\end{align}
which is the same as the one-loop contribution to the Ward identity for that choice. 

While for simplicity we focused on the case with only incoming hard particles, it is possible to generalize our results to  generic in- and out-states by essentially using crossing symmetry. In order to check diagrammatically that crossing is satisfied for dressed states, one can argue that the non-Abelian gluon clouds at leading logarithmic order weakly commute with the S-matrix. The steps are essentially the same as in \cite{Choi:2017ylo}, with the important difference that here we have non-Abelian generators inside the clouds. However even in the non-Abelian theory the coherent state-operators have the property 
\begin{align}
[~\mathcal{U}^{pE}_{\alpha \beta},~ \mathcal{U}^{p^{\prime}E}_{\alpha^{\prime} \beta^{\prime}}] = 0~.
\end{align}
We have explicitly checked that for the ordering prescription $(\text{8.b.2})$ there are no corrections to the Ward identity for generic states, while for ordering $(\text{8.b.1})$ the correction term now has a sum over both outgoing and incoming partons with a corresponding sign. 


\section{Discussion and Conclusions} 
\label{sec:disc}
In this work  we have computed matrix elements of the commutator 
\<
[Q_\epsilon^{\text{lin}}, S]=Q_\epsilon^{+,\text{lin}} S -SQ_\epsilon^{-,\text{lin}}
\>
between asymptotic states corresponding to hard partons dressed with soft-gluon coherent operators. At tree-level the result following from the soft-theorem is essentially the same as in the Abelian case
\begin{align}
&\langle\!\langle  \text{out} \|[Q_\epsilon^{\text{lin}}, S]\| \text{in} \rangle\!\rangle=-\Big[\sum_{\ell\in \text{out}} N^a(\epsilon, p_\ell) t^a_\ell -\sum_{\ell\in \text{in}} N^a(\epsilon, p_\ell) t^a_\ell \Big]~
\langle\!\langle  \text{out} \| S\| \text{in} \rangle\!\rangle~.
\end{align}
In the Abelian case the right-hand side of this relation corresponds to the contribution from the hard part of the asymptotic charge and so this expression is equivalent to the conservation law for the full charge. The non-Abelian theory is quite different as the correct definition of the non-linear charge is unclear and these  difficulties are likely only exacerbated by including quantum corrections. We choose a pragmatic approach of  defining our quantum non-linear hard charge by means of the soft evolution operators and then computing matrix elements to check whether the resulting Ward identities continue to hold. We have seen in \secref{sec:LGTSEv} that the hard charge does receive modifications at loop order however they involve soft gluon operators which can be shown to not contribute at one-loop order if we use IR-finite S-matrix elements with a particular ordering prescription. 

In the computation of matrix elements of the commutator we noted that there are several order-of-limits issues. First, we have assumed a ``weak" definition of the soft charge insertion, where the soft limit for the gluon operator in the charge is taken after the evaluation of the matrix element.  This makes the connection between the asymptotic charge and the standard soft limit of amplitudes most direct. A similar issue arose in the case of sub-leading soft-theorems for graviton amplitudes and our approach is analogous to  that of \cite{Bern:2014oka}.
For the case of the dressed states there remains a second ordering ambiguity, which can be seen in diagrams $(8.\text{b}.1)$ and $(8.\text{b}.2)$ in \figref{fig:OneLoopDSoft}, as one must choose whether to take the soft-limit in the charge before---ordering $\text{O}_1$---or after---ordering $\text{O}_2$---the contribution from the coherent state operator.

These two orderings imply:
\begin{itemize}
	\item Ordering $\text{O}_1$: The orthogonality relation
	\begin{align}
	&\left. \langle\!\langle 0\| Q_\epsilon^{\text{lin}} S \|{\text{in}} \rangle\!\rangle \right|_{\mathcal{O}(\cg^3)}=-
	\frac{\cg^2 C_A}{16\pi^2\hat \epsilon^2} \sum_{\ell\in \text{in}} Q_\epsilon^{\text{h}}(p_\ell)
	\langle\!\langle 0\|S\|{\text{in}} \rangle\!\rangle~.
	\label{eqn:ortho_O1}
	\end{align}
	is broken by pure one loop effects (i.e.\ no tree-level iterated terms) and the soft-charge no longer produces states orthogonal to scattering states.  The Ward identity receives the exact same correction and one may attempt to correct the hard charge at one loop in  fashion similar to \cite{He:2017fsb} to preserve the conservation law. 
	\item Ordering $\text{O}_2$: Using this definition the orthogonality relation is still valid at one-loop and leading singularity
	\<
	\left. \langle\!\langle 0\| Q_\epsilon^{\text{lin}} S \|{\text{in}} \rangle\!\rangle \right|_{\mathcal{O}(\cg^3)}&=&0+\mathcal{O}(\tfrac{1}{\hat \epsilon})
	\>
	and moreover the Ward identity is preserved
	\<
	\left. \langle\!\langle \text{out} \|[Q_\epsilon^{\text{lin}},S] \|\text{in}\rangle\!\rangle \right|_{\mathcal{O}(\cg^3 )}&=&0+\mathcal{O}(\tfrac{1}{\hat \epsilon})~.
	\>
	In this case the interpretation is clearer and we can see that loop effects at this order do not affect the asymptotic symmetries of the S-matrix. 
\end{itemize}
It is quite natural to choose a prescription that preserves the symmetries of the theory where it does not lead to inconsistencies which argues in favor of the second choice. That this second choice is to be preferred on symmetry grounds can be further seen as follows. 
In this work we have used the dressing following from the evolution operator \eqref{eq:softev}. However one could, following \cite{Kulish:1970ut}, modify the dressing so that
\<
\mathcal{J}^a_{q, f}{}_\mu= \cg  \int_{\omega_q} \widetilde{dp}~ \rho^a(\textbf{p})\left(\frac{p_\mu}{p\cdot q}-\frac{f_\mu}{\omega_q}\right)~,
\>
where $f_\mu$ is a $q$-dependent vector and whose value could be determined by insisting on physical asymptotic states. Relatedly one can choose a modified vacuum
\<
\|0\rangle\!\rangle_\Lambda=P_\omega \text{exp}\left(\int^E_\lambda \widetilde{dq} \Lambda^a(q)\Pi^a(q) \right)|0 \rangle
\>
which is an eigenstate of the linearised charge
\<
Q^{ \text{lin} }_\epsilon \| 0 \rangle\!\rangle_\Lambda=\bar{\Lambda}_\epsilon\| 0 \rangle\!\rangle_\Lambda
\>
with 
\<
\bar{\Lambda}_\epsilon=-4\pi \int \widetilde{dq} ~\frac{ \delta(\omega_q) }{ \sqrt{\gamma_{z\bar z} } } \Lambda^a (q)
 (\varepsilon^- \partial_z \epsilon^a  + \varepsilon^+ \partial_{\bar{z}} \epsilon^a)
\>
if we also assume that the Fock vacuum is an eigenstate of the linearised charge with eigenvalue $0$. Strictly speaking this corresponds to the identification of the Fock vacuum 
with the dressed vacuum $|0\rangle\equiv \| 0 \rangle\!\rangle_0$ but this is in fact what we have used as we have set $\Omega^E|0\rangle=|0\rangle$ in our regularised, perturbative calculations. In this sense we may interpret the failure of the orthogonality relation in the case of ordering $\text{O}_1$ as being related to the coherent states no longer being eigenstates of the linearised charge at one-loop and so requiring a one-loop modification. Thus if we make the physically sensible choice that the asymptotic states remain eigenstates of the linearised asymptotic charges we are forced to use ordering $\text{O}_2$ and we thus find that there are no corrections to the Ward identity. 

While our explicit computations have been at one-loop, the leading IR divergence coherent state construction is valid to all-loop orders and so with the correct ordering prescription there should be no leading IR divergent quantum corrections to the Ward identity at any loop order. More non-trivially it should also be possible to repeat the analysis at subleading order in the IR divergences.  The construction of the dressed states is more complicated since single-parton coherent states can no longer be defined and one needs non-factorisable coherent states which take into account two-parton correlations \cite{CATANI1986}.  At sub-leading order the issue of collinear IR-divergences will also become more involved.  In this work we have included collinear effects to the order we consider, that is in the soft-region, which gives rise to the double pole in $\hat \epsilon$, and we see that the asymptotic charges continue to be conserved with the appropriate ordering prescription. At subleading order this will need to be reconsidered as there are also collinear singularities in the non-soft region\cite{Giele:1991vf,Sterman:1995fz,Feige:2014wja}. Such collinear divergences play an important role in understanding tree-level asymptotic symmetries where they are related to the algebra of charges, \cite{Strominger:2013lka, He:2015zea, McLoughlin:2016uwa}. Though at loop-level they are less well studied, the coherent state approach can treat such divergences, see \cite{DelDuca:1989jt, Forde:2003jt} and the recent work \cite{Hannesdottir:2019rqq, Hannesdottir:2019opa} which proposed using the soft-collinear effective theory Hamiltonian to generate the asymptotic time evolution. It can in fact be shown quite generally \cite{GIAVARINI1988} that the dressed S-matrix is completely IR finite and as the proof makes no use of an explicit expression of the asymptotic Hamiltonian it includes any collinear divergences. 

The Faddeev-Kulish dressing of asymptotic particles with soft gauge boson clouds is related to the dressing of hard external particle fields with Wilson lines. This method was first introduced by Mandelstam \cite{Mandelstam:1962mi} in order to achieve a gauge-invariant formulation of QED. Nowadays, it is, amongst other applications, an important tool to efficiently compute the infrared-divergence structure of scattering amplitudes. See e.g.\ \cite{Gardi:2013ita} for a discussion of the virtual soft gluon contribution to multi-leg amplitudes in non-Abelian gauge theories obtained in this formalism. It is also known that the Wilson line approach to soft radiation produces the Faddeev-Kulish dressing when taking the Wilson lines along time-like paths \cite{Jakob:1990zi}. Furthermore, it was recently used in \cite{Choi:2018oel} to study soft photon hair of black holes. It would be interesting to further study the connection between these formulations and asymptotic symmetries in the context of QCD. 

Finally in this work we have focused on single insertions of the asymptotic charge but it would be of interest to consider the generalisation to multiple insertions to study the algebra of asymptotic charges.  He \textit{et al}  in  \cite{He:2015zea}, see also \cite{Cheung:2016iub, Nande:2017dba}, showed that the tree-level double-soft limit of two positive helicity gluons could be rewritten as the level zero Kac-Moody algebra.  This can in fact even be used to construct a stress-energy tensor for gluons by the Sugawara method and to derive a KZ-like equation for MHV amplitudes \cite{McLoughlin:2016uwa}. By considering more general tree-level double-soft limits \cite{Klose:2015xoa, Volovich:2015yoa} the algebra of currents can be extended to include mixed-helicity gluons and sub-leading, in the soft expansion, currents \cite{McLoughlin:2016uwa}. The coherent state approach may make it possible to understand quantum corrections to the algebra of such currents. 

\begin{acknowledgments}
We would like to thank  Einan Gardi, Mario Greco, Lorenzo Magnea, Dhiritman Nandan and Gim Seng Ng for useful discussions. We thank Dhritiman Nandan for comments on an draft version of the current work. This project has received funding from the European Union's Horizon 2020 research and innovation programme under the Marie Sk\l{}odowska-Curie grant agreement No. 764850 "SAGEX". T.McL. and A.S. are supported by the SFI grant 15/CDA/3472. D.M. acknowledges support from the Spanish Government grant FPA2016-78022-P and Spanish MINECO Centro de Excelencia Severo Ochoa Programme (SEV-2016-0597).
\end{acknowledgments}

\appendix

\section{Notations}
\label{sec:appendixA0}

\begin{itemize}
	\item We use a mostly positive signature $g_{\mu\nu}=\text{diag}(-1, +1, +1, +1)$ so that $x^0=ct$ and $p^0=E$.
	\item Retarded coordinates are defined as
	\<
	x^\mu=\left(u+r, r\frac{(z+\bar z)}{1+ z \bar z}, ir\frac{(z-\bar z)}{1+ z \bar z}, r\frac{1-z \bar z}{1+ z \bar z}\right)
	\>
	and polarizations are 
	\<
	\varepsilon^\mu _-=\frac{1}{\sqrt{2}}(z, 1, i , -z)~,~~~\varepsilon^\mu _+=\frac{1}{\sqrt{2}}(\bar{z}, 1, -i , -\bar{z})~.
\>	
	 The polarisation vectors satisfy
\<
\label{eq:pol_rel}
\sum_\sigma \bar{\varepsilon}_\mu^\sigma(q){\varepsilon}_\nu^\sigma(q) =  \eta_{\mu \nu}+c_\mu q_\nu+c_\nu q_\mu~,
\>
where $c_\mu$ is a fixed vector which depends on the choice of polarisation vectors. In our conventions for retarded Bondi coordinates we have $c_\mu=\tfrac{(1+z \bar z)}{2\omega}(-1,0,0,1)$.
	
	\item 
	The free mode expansion of the gluon field is 
	\<
	\mathcal A^a_\mu(x)= \int \widetilde{dq} \big[\bar{\varepsilon}_\mu^\sigma(\vec{q}) a_\sigma^{a}(\vec{q})e^{iq\cdot x}+{\varepsilon}_\mu^\sigma(\vec{q})
	a_\sigma^{a\dagger }(\vec{q}) e^{-iq\cdot x}\big]~,
	\>
	where 
	\<
	\widetilde{dq}=\frac{d^3 q}{(2\pi)^3 (2\omega)}
	\>
	and we use the commutator 
	\<
	[a_\sigma^a(\vec{q}), a_{\sigma'}^{\dagger b}(\vec{q}')]=\tilde{\delta}(\vec{q}-\vec{q}')\delta_{\sigma\sigma'}\delta^{ab}
\>
with $\tilde{\delta}(\vec{q}-\vec{q}')=(2\pi)^3 (2\omega)\delta^{(3)}(\vec{q}-\vec{q}')$.
\item The gauge generators of $\mathfrak{su}(N)$ are defined by the relations
\<
[t^a, t^b]=i f^{abc}t^c
\>
and are normalised such that $\text{Tr}(t^at^b)=\tfrac{1}{2}\delta^{ab}$. Quarks transform in the fundamental representation $t^a=T^a$ and gluons in the adjoint $(t^a)_{b c} = (T_A^a)_{b c}=-i f^a{}_{bc}$.
	\item The covariant derivative is defined as
	\< D_\mu=\partial_\mu-ig_\text{\tiny YM}[\mathcal{A}_\mu,\quad].\>

\end{itemize}
\section{Asymptotic Expansions}
\label{sec:appendixA}
In this appendix we review the construction of the asymptotic charges for non-Abelian gauge theory. We follow \cite{Strominger:2013lka, He:2014cra, He:2015zea, Campiglia:2015qka, Himwich:2019dug} and the review \cite{Strominger:2017zoo}. 
In order to understand the asymptotic symmetries, one must impose fall-off conditions compatible with the equations of motion and which allow for relevant solutions. We consider Lorenz gauge, which in our coordinates is
\begin{align}
-\partial_u (r^2 \mathcal{A}_r) - \partial_r (r^2 \mathcal{A}_u - r^2 \mathcal{A}_r) + \gamma^{z \bar z} (\partial_z \mathcal{A}_{\bar z} + \partial_{\bar z} \mathcal{A}_{z}) = 0
\label{eqn:gauge}
\end{align} 
and impose the conditions at null infinity ($r, t\to \infty$, $u=t-r=\text{const}$)
\<
& &\mathcal{A}_{z}=A_z(u, z, \bar z)+\mathcal{O}(r^{-1})~,~~~\mathcal{A}_{\bar{z}}= A_{\bar z}(u, z, \bar z)+\mathcal{O}(r^{-1})\nn\\
& &\mathcal{A}_{r}=\frac{1}{r^2}A_r(u, z, \bar z)+\mathcal{O}(r^{-3})~,~~~\mathcal{A}_{u}=\frac{1}{r}A_u(u, z, \bar z)+\mathcal{O}(r^{-2})~.
\>
The corresponding condition on the field strength component is 
\<
\mathcal{F}_{ru}=r^{-2} F_{ru}+\mathcal{O}(r^{-3}) ~~~\text{with}~~~~F_{ru}=-(A_u+\partial_u A_r)
\>
which using the $u$-component of the field equations, satisfies
\begin{align}
\partial_u(A_u+\partial_u A_r)- \gamma^{z\bar z} &\partial_u(\partial_z A_{\bar z}+\partial_{\bar z}A_{z}) = -i \cg \gamma^{z\bar z}([A_{\bar z}, \partial_u A_z]+[A_z, \partial_u A_{\bar z}])+\cg j_u^{(2)}~,
\end{align}
where $ j_u=j_u^{(2)}/r^2+\mathcal{O}(r^{-3})$. Hence one finds
\<
Q_\epsilon=\int_{\mathcal{I}^+_-}r^2 d^2 z \, \gamma_{z\bar z}\epsilon^a \mathcal{F}^a_{ru}
\>
giving the expressions \eqref{eq:Q_lin} and \eqref{eq:Q_non-lin}. 

Alternatively, one can impose the following conditions on the falloffs of the non-Abelian gauge fields at large-$r$ \cite{Himwich:2019qmj,Campiglia:2016hvg}
\begin{align}
\mathcal{A}_u&=\frac{1}{r^2} A_u^{(2)}(u,z,\bar z)+\frac{\log(r)}{r}\tilde{A}_u^{(1)}(u,z,\bar z) +\frac{\log(r)}{r^2}\tilde{A}_u^{(2)}(u,z,\bar z) +\mathcal O\left(\frac{\log(r)}{r^3}\right)~,\nonumber\\[0.3cm]
\mathcal{A}_r&=\frac{1}{r^2}A_r^{(2)}(u,z,\bar z)+\frac{\log(r)}{r^2} \tilde{A}_r^{(2)}(u,z,\bar z) +\mathcal O\left(\frac{\log(r)}{r^3}\right)~,\nonumber\\[0.3cm]
\mathcal{A}_z&=A_z^{(0)}(u,z,\bar z)+\frac{1}{r}A_z^{(1)}(u,z,\bar z)+\frac{\log(r)}{r} \tilde{A}_z^{(1)}(u,z,\bar z)+\mathcal O\left(\frac{\log(r)}{r^2}\right)~,
\label{eqn:falloffs}
\end{align}
and use the freedom in the residual gauge transformations at order $\frac{1}{r}$ to set $A_u(u,z,\bar z) = 0$. 
From \eqref{eqn:falloffs} we get the leading components for the field strengths
\begin{align}
\mathcal{F}_{ur}^{a}&= \frac{1}{r^2} \left[\partial_u A_r^{a,(2)} - \tilde{A}_u^{a,(1)}\right] +\frac{\log(r)}{r^2} \left[\partial_u \tilde{A}^{a,(2)}_r + \tilde{A}^{a,(1)}_u \right] + \mathcal O\left(\frac{\log(r)}{r^3}\right)\nonumber \\[0.3cm]
\mathcal{F}_{uz}^{a}&=\partial_u A_z^{a,(0)}+\mathcal O\left(\frac{1}{r}\right)~, \nonumber \\[0.3cm]
\mathcal{F}_{rz}^{a}&= \frac{1}{r^2} \left[-A_z^{a,(2)} + \tilde{A}_z^{a,(1)} - \partial_z A^{a,(2)}_r + g_\text{\tiny YM} f^{a b c} A_r^{b,(2)} A_z^{c,(0)}\right]\nonumber \\[0.3cm]
&+\frac{\log(r)}{r^2} \left(-\tilde{A}^{a,(1)}_z - \partial_z \tilde{A}^{a,(2)}_r + f^{a b c} \tilde{A}^{b,(2)}_r A_z^{c,(0)} \right) +\mathcal O\left(\frac{\log(r)}{r^3}\right)~,  \nonumber \\[0.3cm]
\mathcal{F}_{z\bar z}^{a}&=\partial_zA_{\bar z}^{a,(0)}-\partial_{\bar z}A_z^{a,(0)}+ g_\text{\tiny YM}f^{a b c} A_z^{b,(0)} A_{\bar z}^{c,(0)} +\mathcal O\left(\frac{1}{r}\right)~,
\end{align}
so that the radiation flux is non-zero and finite on $\mathcal{I}^{+}$ as required (\cite{He:2014cra}).
The equations of motion imply the following constraint equations on $\mathcal I^+$ at $\mathcal{O}(1)$ in the large-$r$ expansion
\begin{align}
\label{eqn:eom_Lorenz_gauge}
&-\partial_u \tilde{A}^{a,(1)}_u + \partial_u^2 A_r^{a,(2)} \nn \\
&= - \gamma^{z \bar{z}} \Big[\partial_u( \partial_{\bar{z}} A^{a,(0)}_z + \partial_z A^{a,(0)}_{\bar{z}} )+ g_\text{\tiny YM} f^{a b c} \left(A_{\bar{z}}^{b,(0)} \partial_u A_{z}^{c,(0)} - A_{z}^{c,(0)} \partial_u A_{\bar{z}}^{b,(0)} \right) \Big]+\cg  j_u^{a,(2)} \nonumber
\end{align}
and 
\begin{align}
& 2 (\partial_u A_z^{a,(1)} - \partial_u \tilde{A}_z^{a,(1)}) + \partial_z (\partial_u A_r^{a,(2)} + \tilde{A}_u^{a,(1)}) - g_\text{\tiny YM} f^{a b c} \partial_u (A_r^{b,(2)} A_z^{c,(0)})- g_\text{\tiny YM} f^{a b c} \tilde{A}_u^{b,(1)} A_z^{c,(0)}  \nonumber \\
&- \partial_z \left(\gamma^{z \bar{z}} (\partial_z A^{a,(0)}_{\bar{z}} - \partial_{\bar{z}} A^{a,(0)}_{z} + g_\text{\tiny YM} f^{a b c} \tilde{A}_z^{b,(0)} A_{\bar{z}}^{c,(0)})\right)\nn\\
&+g_\text{\tiny YM} f^{a b c}\Big[-A_r^{b,(2)} \partial_u A_z^{c,(0)} + \gamma^{z \bar{z}} A_z^{b,(0)} \Big(\partial_{\bar{z}} A_z^{c,(0)} - \partial_{z} A_{\bar{z}}^{c,(0)} + g_\text{\tiny YM} f^{c d e} \tilde{A}_{\bar{z}}^{d,(0)} A_{z}^{e,(0)})\Big) \Big]=\cg  j_z^{a,(2)} 
\end{align}
and
\begin{align}
&-\partial_u A_r^{a,(2)} - \tilde{A}_u^{a,(1)} + \gamma^{z \bar{z}} (\partial_z A_{\bar{z}}^{a,(0)} + \partial_{\bar{z}} A_{z}^{a,(0)}) = 0
\end{align}
whereas at order $\mathcal{O}(\log(r))$
\begin{align}
\label{eqn:log_modes}
& 2 \partial_u \tilde{A}^{a,(1)}_z - 2 \partial_z \tilde{A}^{a,(1)}_u - g_\text{\tiny YM} f^{a b c} \Big(\partial_u \tilde{A}_r^{b,(2)} A_z^{c,(0)} + \tilde{A}^{b,(1)}_u A_z^{c,(0)} +\tilde{A}^{b,(2)}_u A_z^{c,(0)} + \tilde{A}^{b,(2)}_r \partial_u A_z^{c,(0)} \Big)  = 0.
\end{align}
In these expressions we have used that the currents have the decay properties
\begin{align}
\label{eqn:decay_current}
j_u = \mathcal{O}\left(\frac{1}{r^2}\right) \qquad j_r = \mathcal{O}\left(\frac{1}{r^3}\right) \qquad j_z , j_{\bar{z}} = \mathcal{O}\left(\frac{1}{r^2}\right)~.
\end{align}
Just as before \eqref{eqn:eom_Lorenz_gauge} can be used to rewrite the charge as in \eqref{eq:Q_lin} and \eqref{eq:Q_non-lin}. One can also use retarded radial gauge or radiation gauge and at least at leading order one ultimately finds the same expressions
for the charge. 
%

\nocite{*}

\bibliographystyle{JHEP-2}
\bibliography{references}

\end{document}